\documentclass[10pt, twocolumn, comsoc]{IEEEtran}

\usepackage{graphicx,epsfig}
\usepackage[noadjust]{cite}
\usepackage{mcite}
\usepackage{amsfonts,helvet}
\usepackage{fancyhdr}
\usepackage{threeparttable}
\usepackage{epsf,epsfig}
\usepackage{amsthm}
\usepackage{amsmath}
\usepackage{boldline}
\usepackage{booktabs}
\usepackage{amssymb}
\usepackage{lipsum}
\usepackage[caption=false,font=footnotesize]{subfig}
\usepackage{textcomp}

\usepackage{ragged2e} 
\usepackage{multirow} 
\usepackage{makecell} 

\usepackage[colorlinks=true, linkcolor=blue]{hyperref}

\usepackage{dsfont}
\usepackage{color}
\usepackage{array}
\usepackage{algpseudocode}
\usepackage{algcompatible}
\usepackage{enumerate}
\usepackage{gensymb}
\usepackage{cancel}

\usepackage[linesnumbered,ruled,noend]{algorithm2e}

\usepackage{fancyhdr}
\usepackage{graphicx,epsfig}
\usepackage{wrapfig}
\usepackage{ragged2e}
\usepackage{algpseudocode}
\usepackage{algcompatible}
\usepackage{threeparttable}

\newtheorem{theorem}{Theorem}

\newtheorem{corollary}{Corollary}

\newtheorem{lemma}{Lemma}

\newtheorem{proposition}{Proposition}

\usepackage{eucal}

\begin{document}

\title{ Spherical Point Process with Random Heights: \\New Approach for Modeling and Analysis of Downlink Satellite Networks
}


\author{Seyong~Kim, Jinseok~Choi, Namyoon~Lee, François~Baccelli, and Jeonghun~Park

\thanks{This work was supported by Institute of Information \& communications Technology Planning \& Evaluation (IITP) under 6G Cloud Research and Education Open Hub(IITP-2025-RS-2024-00428780) grant funded by the Korea government(MSIT). S. Kim and J. Park are with the School of Electrical and Electronic Engineering, Yonsei University, Seoul 03722, South Korea (e-mail: {\texttt{sykim@yonsei.ac.kr; jhpark@yonsei.ac.kr}}). J. Choi is with School of Electrical Engineering, KAIST, South Korea (email: {\texttt{jinseok@kaist.ac.kr}}),  N. Lee is with Department of Electrical Engineering, POSTECH, Pohang, South Korea (e-mail: {\texttt{nylee@postech.ac.kr}}). F. Baccelli is with INRIA-ENS, France (email: {\texttt{francois.baccelli@ens.fr}})
}}

\maketitle

\begin{abstract}
The Low Earth Orbit (LEO) satellite industry is undergoing rapid expansion, with operators competitively launching satellites due to the first-come, first-served principle governing orbital rights. This has led to the formation of increasingly large-scale, volumetric constellation where satellites operate across a diverse range of altitudes. To address the need for analyzing such complex networks, this paper establishes a new analytical framework for LEO constellations by leveraging a 3D Poisson point process (PPP). 
Specifically, we introduce a random height model (RHM) that can capture various altitude distributions by applying a random radial displacement to points generated by a homogeneous PPP on a nominal shell. Building on this, we derive an analytical expression for the downlink coverage probability. To motivate our model, we show that the altitude distributions of several leading satellite constellations, including Starlink, align with our model's assumptions. We then demonstrate through Monte Carlo simulations that the coverage probability of our RHM closely matches that of these real-world networks. Finally, we confirm the accuracy of our analytical expressions by showing their agreement with simulation results. Our work thereby provides a powerful tool for understanding and predict how the statistical distribution of satellite altitudes impacts network performance.
\end{abstract} 


\section{Introduction}
Interest in satellite networks for seamless global coverage is accelerating across corporate and national domains, driven by applications from commercial connectivity to national security. 
In particular, 
Low Earth Orbit (LEO) satellites have gained significant attention due to their relatively low transmission delay and high data rate \cite{liu:commmag:21}. 
However, unlike geostationary (GEO) satellites \cite{kim:twc:25}, 
LEO satellites move at high orbital velocities, which necessitates deploying a large number of satellites to maintain continuous global coverage. 
This requirement has fueled a surge of large-scale constellation deployments, initially led by SpaceX’s Starlink \cite{mcdowell:aas:20} and followed by other operators such as OneWeb and Globalstar. 
For instance, Starlink has received FCC approval for 7,500 second-generation satellites \cite{spacex:fcc:22} and is awaiting authorization for an additional 22,488 satellites. Furthermore, recent regulatory approvals have expanded Starlink’s operational altitudes to new orbital shells between 340 km and 360 km \cite{spacex:fcc:24}, complementing its existing operations around 500 km.


As LEO constellations continue to expand in both scale and architectural complexity, 
the demand for rigorous yet tractable network analysis is becoming increasingly critical in both industry and academia. Although sophisticated simulation tools such as the Systems Tool Kit (STK) and the General Mission Analysis Tool (GMAT) provide high-fidelity orbital dynamics modeling, they are computationally intensive and less suitable for deriving generalizable system-level insights. This highlights the need for analytical frameworks that balance realism and tractability. 



To address this need, recent works have applied stochastic geometry to the modeling of satellite networks. Originally developed for modeling terrestrial cellular networks on a 2D plane, stochastic geometry has been extended to capture the spatial characteristics of 3D satellite networks. 
A common approach in existing studies is to place a point process on a spherical shell, where binomial \cite{okati:tcom:20}, Poisson \cite{park:twc:22}, or Cox \cite{choi:tcom:2025} processes are used to represent satellite locations. While such models offer analytical tractability, they are limited in capturing altitude variations among satellites. 
In this paper, we put forth a new approach to modeling satellite networks. Specifically, we construct a Poisson point process on the sphere and assign to each point an independent altitude mark, yielding a tractable framework that naturally captures random satellite altitudes. 
We demonstrate that this framework achieves analytical tractability while accurately representing the spatial characteristics of realistic satellite deployments. 




\begin{figure*}[t]
  \centering
  \subfloat[]{\includegraphics[width=0.4\textwidth]{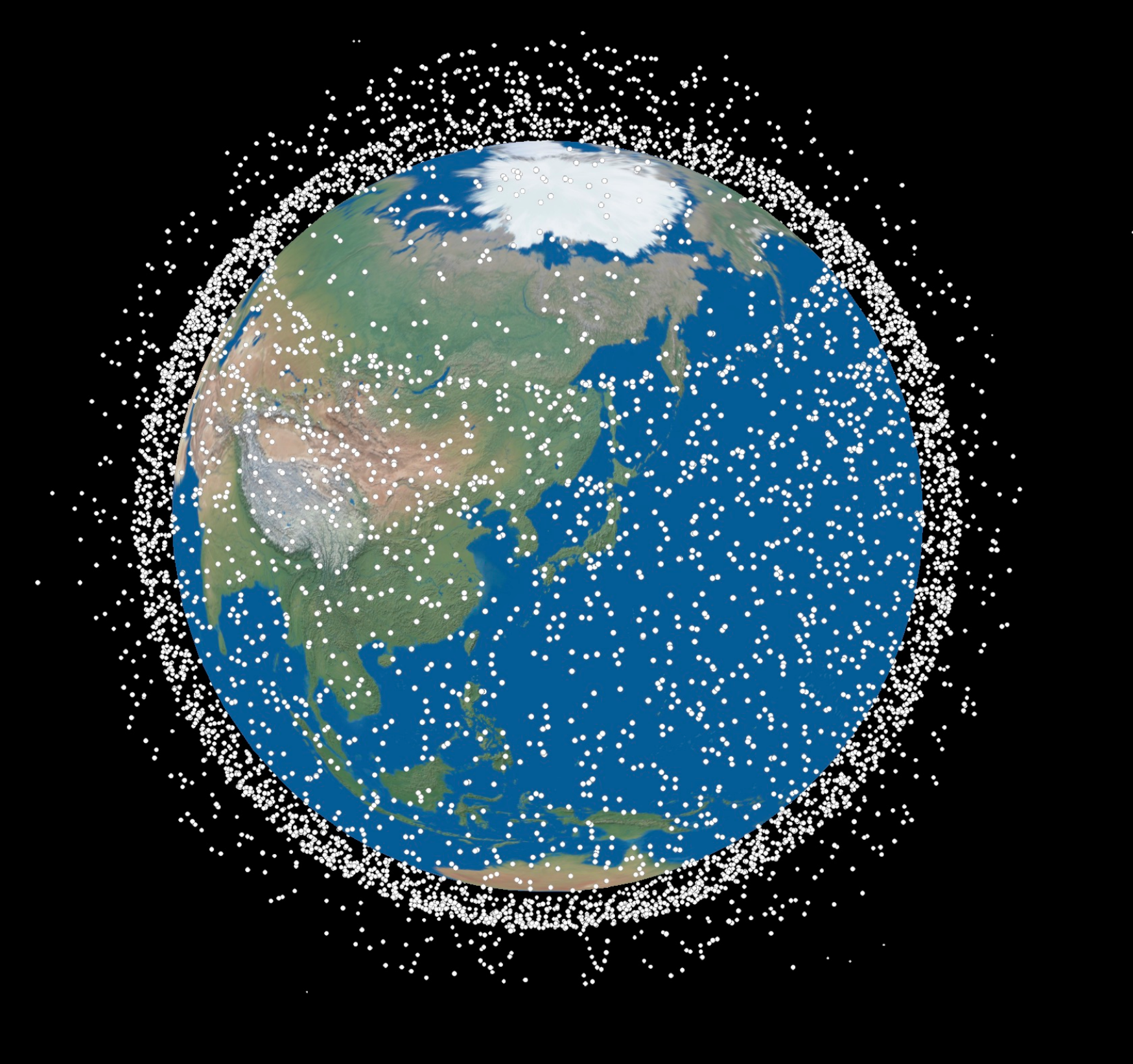}}
  \hfill
  \subfloat[]{\includegraphics[width=0.55\textwidth]{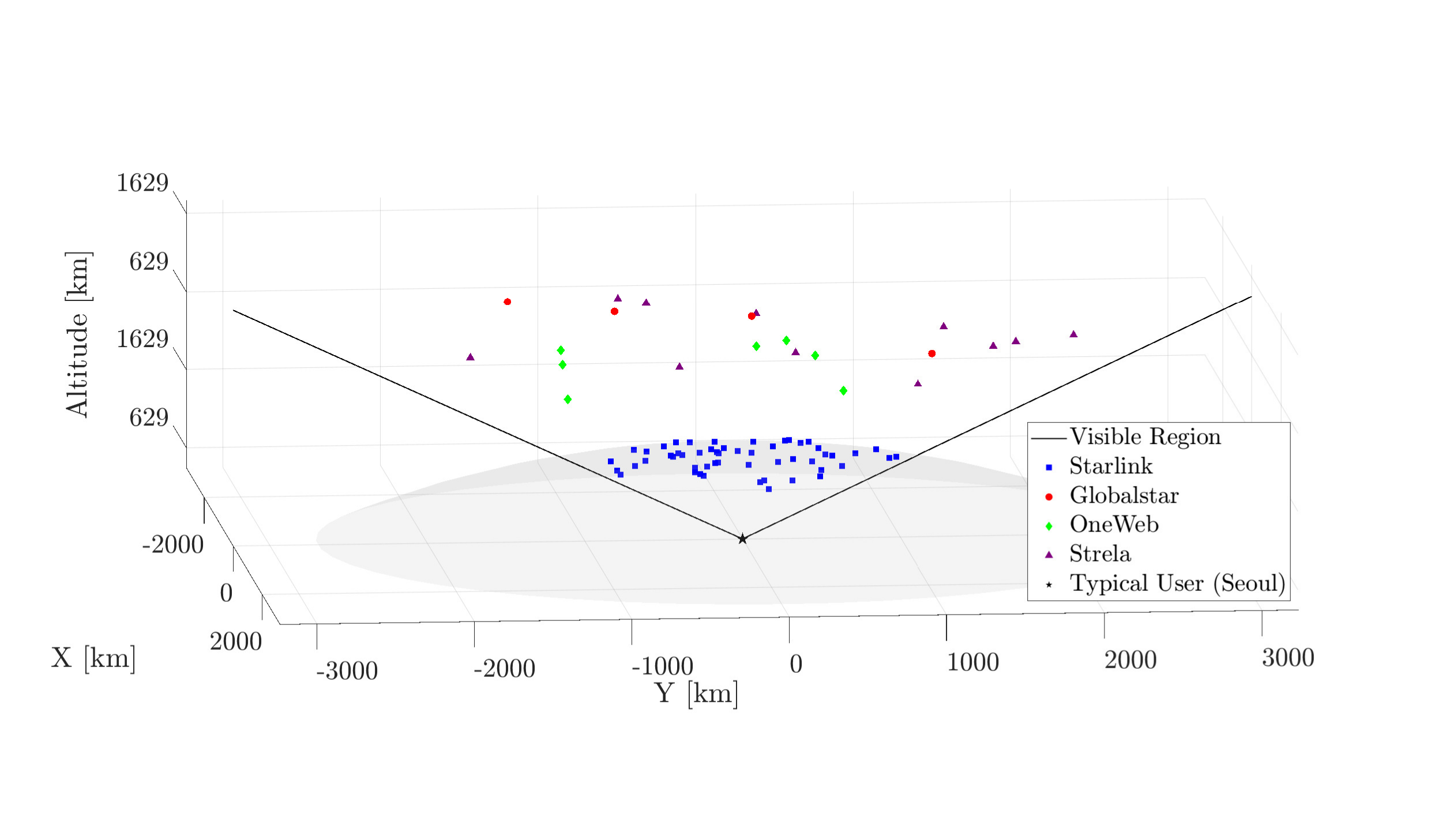}}
  \caption{The multi-altitude distribution of four major commercial constellations: Starlink, OneWeb, Strela, and Globalstar. (a) The complete global distribution of the constellations (Source: https://satellitetracker3d.com), and (b) the resulting subset of satellites visible to a typical ground user (Source: https://N2YO.com).}
  \label{fig:intro figure}
\end{figure*}

\subsection{Motivation and Related Works} 
Stochastic geometry has been widely employed as a mathematical tool to analyze system-level performance of wireless networks, where the spatial locations of base stations and users are modeled by a point process \cite{baccelli:book:09, andrews:tcom:11, park:twc:16}. 
In the context of 2D terrestrial cellular networks, this approach offers intuitive modeling and tractable performance analysis, which has provided valuable network insights and spurred active research \cite{andrews:tcom:11}. This application was further extended to other complex network models such as multilayer \cite{dhillon:jsac:12, renzo:twc:15, park:twc:18, park:tccn:18} and unmanned aerial vehicle (UAV) networks \cite{tian:twc:22, hou:tcom:2019, lahmeri:commlett:2020, hayajneh:access:2018}. 
Subsequently, this research direction has moved towards 3D non-terrestrial networks (NTNs). 
A comprehensive overview of NTN modeling using stochastic geometry is detailed in \cite{wang:surv:2025}. 

As a major component of NTNs, satellite networks are a key enabler for ubiquitous global connectivity, driving intense research into their system-level modeling. 
In early studies, a homogeneous binomial point process (BPP) was a natural modeling choice \cite{talgat:commlett:20, okati:tcom:20, okati:pimrc31:2020, okati:commlett:2023, ma:pimrc35:2024} as it can directly reflects the fixed number of satellites in actual constellations. 
Specifically, in \cite{okati:tcom:20}, the authors modeled satellite locations as a BPP on a spherical shell with a fixed height and derived the typical user's downlink coverage probability. Extending this work, \cite{okati:pimrc31:2020} derived both uplink and downlink coverage probabilities in a mega-constellation environment. 

Despite this intuitive appeal, the mathematical complexity and intractable nature of BPP modeling highlighted the need for a more tractable approach, such as the Poisson point process (PPP) model.
In \cite{hourani:wcl:21}, although the total number of satellites was fixed at $N$, the authors approximated the satellite locations as a PPP for analytical tractability, demonstrating that this PPP approximation holds tight even for a small number of satellites $N$. Similarly, \cite{homssi:commlett:2021} approximated the BPP with a PPP to derive the uplink coverage performance for tractability. 
Subsequently, in \cite{park:twc:22}, the satellite distribution was modeled as a PPP at a single-altitude, and the typical user's coverage probability was derived and validated against the actual Starlink constellation. It was confirmed that for a very large $N$, the performance of PPP and BPP models becomes very tight. 


However, conventional BPP and PPP models typically assumed a uniform distribution across the whole spherical region, which does not fully capture the orbit-dependent patterns of actual satellite constellations. 
This modeling deficiency can lead to noticeable deviations in performance evaluation relative to actual constellation behavior.
To resolve this, BPP models introduce an effective number of satellites dependent on latitude to reduce the moment mismatch between the model and the actual distribution \cite{okati:tcom:20, okati:pimrc31:2020}. 
Unfortunately, calculating this requires the probability density function (PDF) of the latitude, which is often complex and intractable. 
Similarly, PPP-based models require parameter tuning to mitigate performance gaps; for example, \cite{park:twc:22} adjusts the satellite density to better align the model with the performance of the Starlink constellation. 
To address this issue more directly, research has moved towards analyzing satellite networks using orbit-based stochastic geometry \cite{choi:tvt:2024, lee:access:2024}. In \cite{choi:tvt:2024}, for instance, the authors applied a Cox point process (Cox) to create an orbit-dependent model of the satellite constellation, analyzing the no-satellite probability and coverage probability. In \cite{lee:access:2024}, satellites were modeled as a PPP on each orbit, and the system's coverage probability was derived by considering the orbital inclination.

While stochastic geometry has been widely applied to the system-level analysis of satellite networks, most studies rely on a critical simplification: modeling satellites at a single, fixed altitude \cite{okati:tcom:20, okati:pimrc31:2020, park:twc:22, choi:tcom:2025, lee:access:2024}. In practice, however, modern satellite deployments are far more complex. For instance, Fig.~\ref{fig:intro figure} (a) provides a snapshot of operational satellite constellations based on real-world tracking data, demonstrating that major services such as Starlink, Globalstar, OneWeb, and Strela are deployed across a broad range of altitudes to support diverse services and ensure robust connectivity. This altitude variation is not arbitrary but a necessity, driven by the operational demands of mega-constellations, including collision avoidance with other satellites and debris.

From a user's perspective, this results in a complex three-dimensional cloud of visible satellites, as shown for a user in Seoul in Fig.~\ref{fig:intro figure} (b). This discrepancy between simplified models and physical reality creates a critical research gap. Although existing research has started to address multi-altitude scenarios through models like the BPP \cite{okati:commlett:2023, ma:pimrc35:2024} and the Cox process \cite{choi:tvt:2024}, these approaches are limited in some aspects. 
They not only require complex, system-specific parameter tuning (i.e., moment matching) but have also been restricted to specific cases, such as a uniform distribution of altitudes \cite{choi:tvt:2024}. 
Strikingly, while the PPP is widely recognized as a more tractable and scalable tool, its application has thus far been confined to single-altitude scenarios, leaving the multi-altitude case unexplored. This paper fills this critical gap by developing the first tractable, PPP-based framework for analyzing satellite networks with arbitrary random altitude distributions.

\subsection{Contributions and Organization} 
In this paper, we propose a tractable framework for analyzing 3D satellite constellations. This approach extends the conventional single-altitude shell model to incorporate random satellite heights. Our model is constructed as follows: we first distribute the satellites according to a PPP on a nominal sphere of radius $R_{\sf S}$, centered at the Earth's origin.
Subsequently, each satellite is radially displaced by a random height $h$, representing its altitude perturbation from the nominal shell. The ground users are modeled as an independent PPP on the Earth's surface, a sphere of radius $R_{\sf E}$.
Leveraging Slivnyak’s theorem, our analysis is conducted for a typical user located at $(0,0,R_{\sf E})$ without loss of generality.
This user is served by the nearest visible satellite, defined as the one with the minimum distance within a visible region determined by an elevation angle $\theta$, while treating all other visible satellites as sources of interference.
Based on this framework, our main contributions are summarized as follows:

\begin{itemize} 
    \item This paper proposes the first tractable analytical framework for volumetric LEO constellations using a PPP-based Random Height Model (RHM), where satellite altitudes are randomly distributed around a nominal value. Within this framework, we derive fundamental performance metrics, including the probability of satellite presence, the nearest satellite distance distribution, the conditional Laplace transform of interference, and ultimately, the downlink coverage probability.
    \item 
    We validate our derived coverage probability by comparing our model's predictions against the performance of commercial LEO satellite constellations. Through this comparison, our analysis reveals a key insight: the performance discrepancy between theoretical models and reality is critically driven by the variance of the altitude distribution, indicating that mean altitude alone is an insufficient indicator of performance. 
    Our findings thus confirm the necessity of the RHM to accurately capture the characteristics of deployed satellite networks. Consequently, we demonstrate that the RHM significantly reduces the mismatch between theoretical predictions and real-world performance. Through extensive simulations, our model provides a much more accurate coverage probability estimate compared to conventional stochastic geometry models, by effectively capturing the impact of altitude variance. 
    \item
    We incorporate a realistic Shadowed-Rician (SR) fading model which is a key differentiator from prior works that used simpler Nakagami fading \cite{park:twc:22, choi:tcom:2025, choi:tvt:2024, okati:commlett:2023} or Rayleigh fading \cite{okati:tcom:20, ma:pimrc35:2024}. To ensure analytical tractability, we then employ a two-step simplification: first, we approximate the SR fading with a highly accurate Gamma distribution. Second, by leveraging Alzer's inequality, we derive a simple and tractable expression for the coverage probability. Our results confirm this final expression is highly accurate, with the marginal error being a worthwhile trade-off for its significant reduction in computational complexity.
\end{itemize} 

The remainder of this paper is organized as follows. Section II details the system model, outlining the network geometry, channel characteristics, and key performance metrics, along with the necessary mathematical preliminaries. Section III presents our main theoretical contribution: the derivation of the coverage probability, yielding a simply tractable expression. In Section IV, we provide extensive numerical results to validate our analytical framework, investigate the impact of various satellite altitude distributions, and demonstrate the accuracy of our findings. Finally, Section V concludes the paper.


\section{System Model}
We explain the RHM in spherical random point process for satellite communications and the performance metric to analyze the coverage probability.

\subsection{Network Model}

\begin{figure}[t]
  \centering
  \subfloat[]{\includegraphics[width=0.235\textwidth]{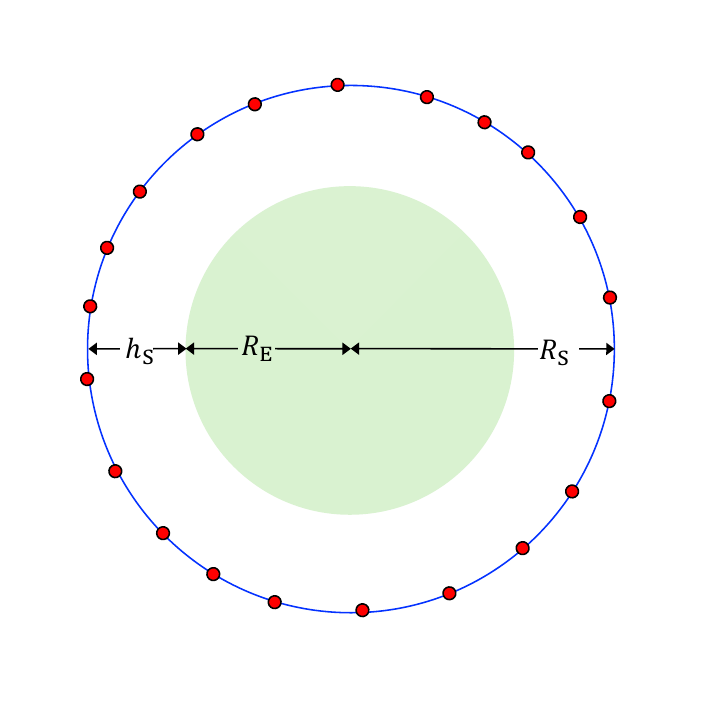}}
  \hfill
  \subfloat[]{\includegraphics[width=0.235\textwidth]{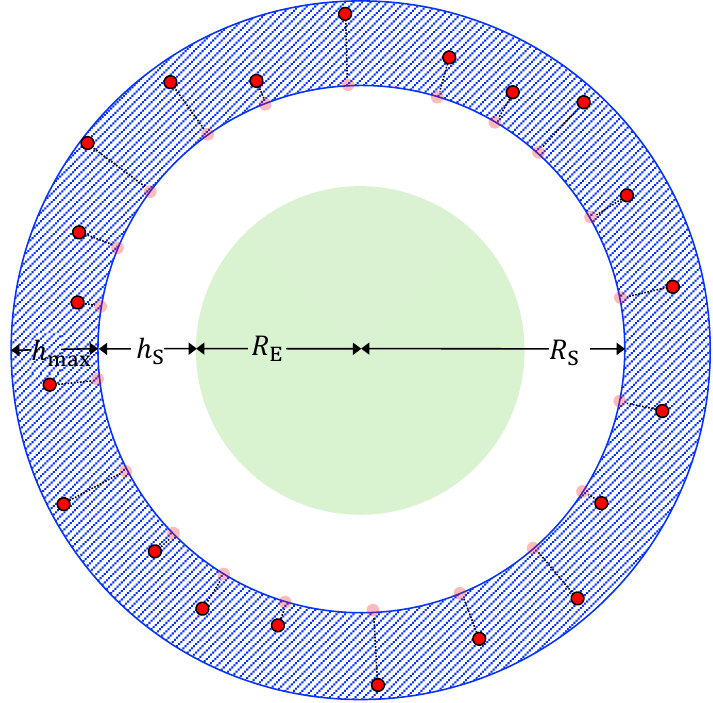}}
  \caption{Illustration of satellite networks. (a) is general SPPP where the points are distributed on the surface of sphere with radius $R_{\sf S}$. (b) is the proposed RHM where each point on the surface of SPPP has random height toward radial direction.}
  \label{fig:RHM}
\end{figure}


{\bf{Poisson point process on a sphere}}: To model the spatial locations of satellite, we consider PPPs distributed on a sphere (SPPP) \cite{park:twc:22}. For better understanding, we first introduce a generic SPPP model with random heights, and then present the specific model for the satellite networks.
Consider a sphere defined in $\mathbb{R}^3$ whose center is the origin ${\bf{0}}$ and radius is fixed as $R$.
Let $\Phi$ be an isotropic PPP on $\mathbb{S}_{R}^2$ where
\begin{align}
    \mathbb{S}_{R}^2=\{{\bf x}\in \mathbb{R}^3: \|{\bf x}\|_2=R\}.
\end{align}
We denote the points of this point process as $\{{\bf x}_i^{}, 1 \le i \le N_{}, \| {\bf x}_i^{} \|_2 = R\}$ and denote the density of $\Phi_{}$ as $\lambda_{}$. The number of points on the sphere $N$ is a random variable drawn from the Poisson distribution with mean $4 \pi R_{}^2 \lambda$. 
The points are assumed to be independently and uniformly distributed on the surface of $\mathbb{S}^2_R$, i.e., we have a homogeneous SPPP.
We denote the surface of $\mathbb{S}^2_R$ as $\mathcal{A}$ while the number of points of $\Phi$ located in a particular set $\mathcal{A} \subset {\mathbb{S}}_R^2$ is denoted as $\Phi(\mathcal{A})$. 

As the deployment density of the satellite network increases, the analytical coverage probability derived under this homogeneous SPPP model and the actual coverage probability computed by using a realistic Starlink constellation set closely match as demonstrated in \cite{park:twc:22}. 

{\bf{SPPP with Random heights}}: 
Based on the SPPP $\Phi$, we define a new point process $\hat{\Phi}$, which is obtained by a random transformation of the original points, an operation formally known as the displacement theorem \cite{baccelli:book:09}. Specifically, each point $\mathbf{x}_i \in \Phi$ is radially displaced by an independent and identically distributed (IID) random height $h_i$, drawn from a distribution $f_H$ with support set $\CMcal{H}$. The new location of each point is:
\begin{align}
    \hat{\Phi} = \left\{ \hat{\mathbf{x}}_i \mid \hat{\mathbf{x}}_i = \mathbf{x}_i \cdot \frac{R+h_i}{R},\;\text{for $\mathbf{x}_i \in \Phi$},\;h_i \sim f_H \right\},
\end{align}
where $\| \hat {\bf{x}}_i \| = R + h_{i}$.
To analyze the properties of $\hat{\Phi}$, we use the fact that this construction is mathematically equivalent to creating a marked point process. In this framework, each point  $\mathbf{x}_i$ is assigned a random mark $h_i$. A key principle of PPPs, known as the superposition (or coloring) theorem, states that if we group the points by their marks, the resulting sub-processes are also independent PPPs. Therefore, the overall process $\hat{\Phi}$ is a superposition of the independent point processes $\{\hat{\Phi}^h\}_{h \in \CMcal{H}}$:
\begin{align}
    \hat{\Phi} = \bigcup_{h \in \CMcal{H}} \hat{\Phi}^h.
\end{align}
In this model, each sub-process $\hat{\Phi}^h$ is itself an independent PPP whose points share the same mark $h$ and are spatially restricted to a unique spherical shell with radius $R+h$.


{\bf{Satellite network}}: 
Building on the general framework, we now define the satellite point process with random heights, which is based on two concentric spheres: the Earth with radius $R_{\sf E}$, and the satellite orbital sphere with radius $R_{\sf S} (>R_{\sf E})$. We consider a SPPP distributed on the surface of $\mathbb{S}_{R_{\sf S}}^2$ with density $\lambda$ as
\begin{align}
    \Phi_{\sf S} = \{\mathbf{d}_i \in \mathbb{S}_{R_{\sf S}}^2, \|\mathbf{d}_i\|^2=R_{\sf S},\; 1 \le i \le N_{}\},
\end{align}
where $N$ follows a Poisson distribution with mean $\lambda 4\pi R^2_{\sf S}$.
Now, we define SPPP with random height by radially displacing $\mathbf{d}_i$ in $\Phi_{\sf S}$ by a random height $h_i$ as
\begin{align}
    \hat{\Phi}_{\sf S} = \left\{ \hat{\mathbf{d}}_i \mid \hat{\mathbf{d}}_i = \mathbf{d}_i \cdot \frac{R_{\sf S}+h_i}{R_{\sf S}},\;\text{for $\mathbf{d}_i \in \Phi_{\sf S}$},\;h_i \sim f_H   \right\}.
\end{align}
As considering two spheres sharing the same origin, we define $h_{\sf S} = R_{\sf S} - R_{\sf E}$, where $h_{\sf S}$ is the standard satellite altitude, as illustrated in Fig.~\ref{fig:RHM} (a). 
We assume that the heights are IID over the interval $\CMcal{H} = [0, h_{\sf{max}}]$ where $h_{\sf{max}} \ge 0$. Thus, the probability density function (PDF) is given by $f_H(h) \sim \mathrm{Unif}(\CMcal{H})$. This illustration of RHM can be seen in Fig.~\ref{fig:RHM} (b). 
Although this paper assumes a uniform distribution for analytical simplicity, the proposed methodology is not restricted to any particular distribution and remains valid for any arbitrary distribution $f_H(h)$.

Let $\hat{\Phi}_{\sf S}$ denote the aggregate process where each satellite is displaced by a random height $h \in \CMcal{H}$. We decompose $\hat{\Phi}_{\sf S}$ into independent marked sub-processes $\hat{\Phi}^h_{\sf S}$, each corresponding to satellites displaced by a specific height $h$:
\begin{align}\label{eq:Phi h}
\hat{\Phi}_{\sf S} = \bigcup_{h \in \CMcal{H}} \hat{\Phi}^h_{\sf S}.
\end{align}

In this modeling, $\hat \Phi_{\sf S}$ incorporates the scenario where satellites are positioned at varying altitudes. If $h_i = 0$ for all $i$, each satellite is located at the fixed altitude $h_{\sf S}$, which coincides with the network model presented in \cite{park:twc:22}.

\begin{figure*}[t]
  \centering
  \subfloat[]{%
    \includegraphics[width=0.235\textwidth]{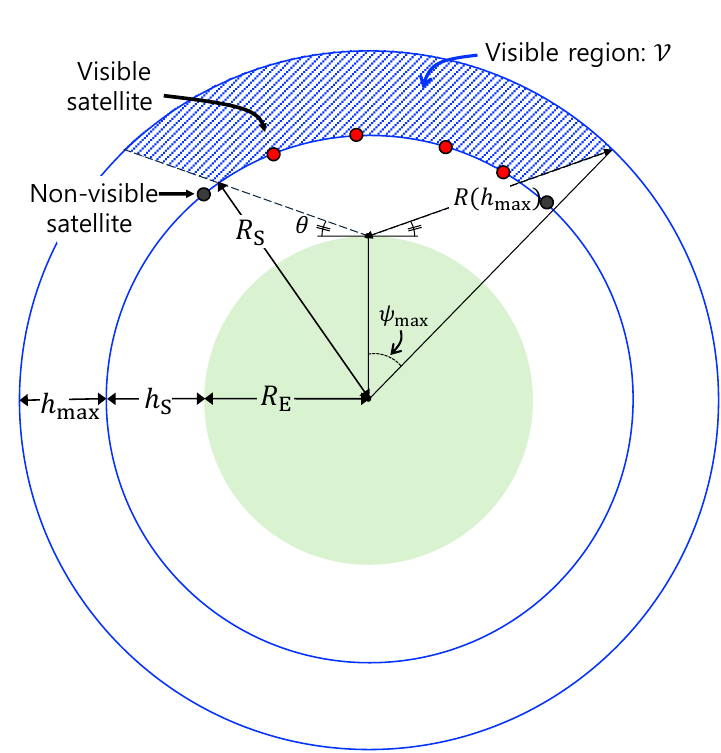}%
    \label{fig:conventional_model}%
  }
  \hfill
  \subfloat[]{%
    \includegraphics[width=0.235\textwidth]{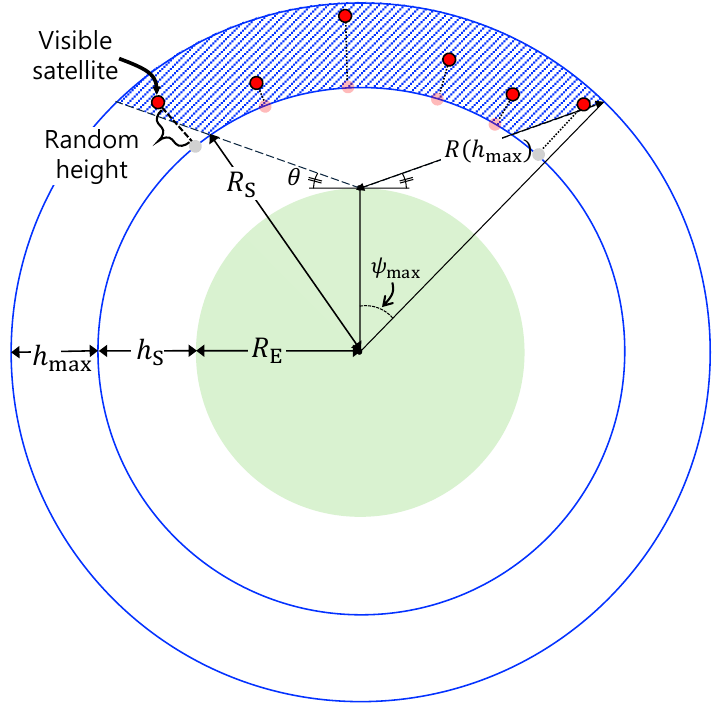}%
    \label{fig:proposed_model}%
  }
  \hfill
  \subfloat[]{%
    \includegraphics[width=0.235\textwidth]{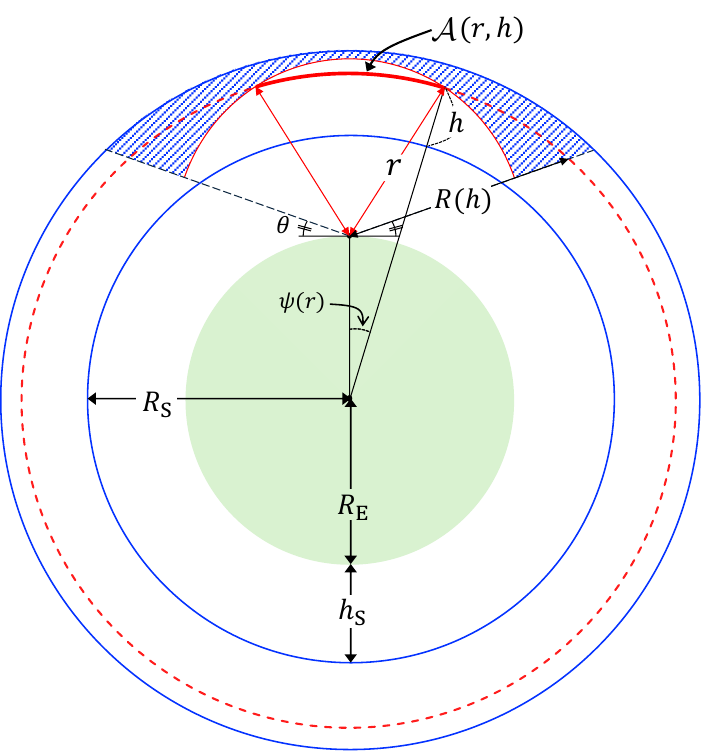}%
    \label{fig:A(r,h)}%
  }
  \hfill
  \subfloat[]{%
    \includegraphics[width=0.235\textwidth]{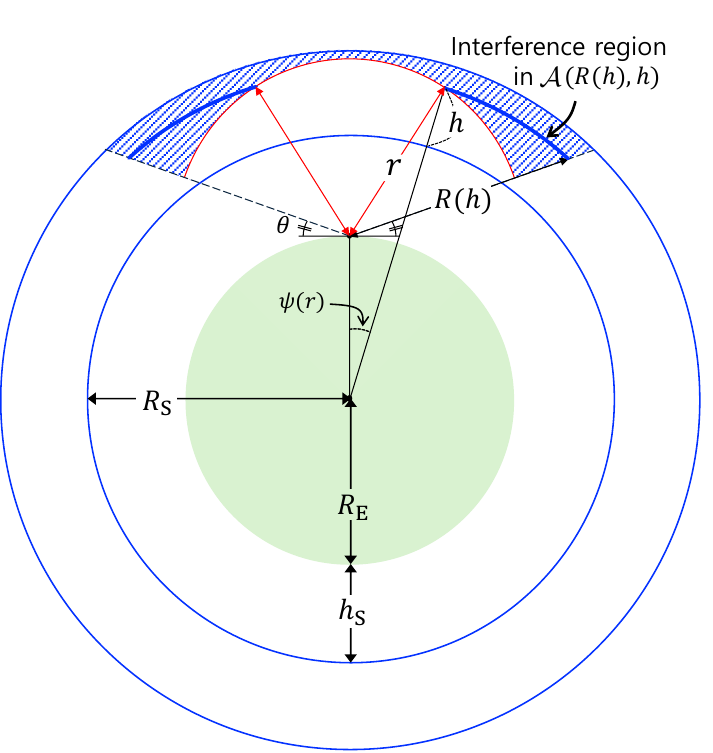}%
    \label{fig:cap laplace}%
  }
  \caption{(a) Illustrates the visible region in a traditional single-altitude model, where all satellites share one altitude, forming a 2D spherical cap. (b) Shows the visible region under the RHM, where satellites at various altitudes expand the region from a 2D cap into a 3D volume. (c) Defines the geometry at altitude $h$: the spherical cap  $\mathcal{A}(r,h)$ is the region within a distance $r$ of the user, and $R(h)$ is the maximum distance defining the total visible area. (d) Depicts the interference region at altitude $h$, defined as the total visible region with a central exclusion zone of radius $r$ removed.}
  \label{fig:system_model}
\end{figure*}

{\textbf{Typical user}}: We also model the users' locations as a homogeneous SPPP with density $\lambda_{\sf U}$ on the Earth's surface, $\mathbb{S}^2_{R_{\text{E}}}$. We denote this point process as $\Phi_{\sf U} = \{{\bf u}_i, 1 \le i \le N_{\sf U}, \| {\bf u}_i \|_2 = R_{\sf E}\}$, where the total number of users $N_{\sf U}$ follows a Poisson distribution with mean $4\pi R^2_{\sf E}\lambda_{\sf U}$. Since the user process $\Phi_{\sf U}$ and the satellite process $\hat \Phi_{\sf S}$ are independent and invariant under rotations in $\mathbb{R}^3$, we can apply Slivnyak's theorem to analyze the system from the perspective of a typical user \cite{baccelli:book:09}. Without loss of generality, we place this user at the Cartesian coordinates $(0, 0, R_{\sf E})$, a choice which does not alter the statistical distribution of the satellite process $\hat \Phi_{\sf S}$.

\textbf{Typical spherical cap}:  
The analysis of satellite visibility, which determines the set of potential signal and interference sources, is more complex under the RHM than in a single-altitude model. In the latter case, as depicted in Fig.~\ref{fig:system_model} (a), all satellites orbit a single sphere, and the visible region is simply a spherical cap. For the RHM, however, varying satellite altitudes expand this region into a three-dimensional volume, as shown in Fig.~\ref{fig:system_model} (b). Therefore, characterizing this visible volume is a critical first step in performing the coverage analysis. 

To do this, we introduce the concept of a spherical cap from the perspective of the typical user, who is located at $\mathbf{u}_1  =(0,0,R_\text{E})$.
We define the typical spherical cap as $\mathcal{A}(r, h)$ as the region on the surface $\mathbb{S}^2_{R_{\text{S}}+h}$ containing all points whose line-of-sight (LoS) distance from the user is no more than $r$, as illustrated in Fig.~\ref{fig:system_model} (c):
\begin{align}
    \mathcal{A}(r,h) = \left\{\hat{\mathbf{d}} \in \mathbb{S}^2_{R_{\text{S}}+h} \middle| \|\hat{\mathbf{d}} - \mathbf{u}_1 \| \le r \right\}.
\end{align}
The area of $\mathcal{A}(R, h)$ is given by
\begin{align}\label{eq:A(R,h)}
    A(R,h) = |\mathcal{A}(R, h)| = \frac{\pi(R_{\sf S}+h)(R^2-(h_{\sf S }+h)^2)}{R_{\sf E}},
\end{align}
where $h_{\sf max}$ is defined such that $h_{\sf S} - h_{\max} > 0$.
The term $R(h)$ represents the maximum possible $r$ for a given $h$, which is determined by the minimum elevation angle $\theta$ as depicted in Fig.~\ref{fig:system_model} (c):
\begin{align}\label{eq:R(h)}
    R(h) = \sqrt{(R_{\sf S} + h)-R^2_{\sf E}\cos^2(\theta)}-R_{\sf E}\sin(\theta).
\end{align}
The distance $r$ of interest, therefore, lies in the range
$R_{\sf min} \le r \le R_{\sf max}$, where
\begin{align}
    R_{\sf min} = h_S - h_{\max}, \;\;
    R_{\sf max} = R(h_{\sf max}).
\end{align}

Next, we define the visible space $\mathcal{V}\subseteq\mathbb{R}^3$ as the three-dimensional region containing all points visible to the typical user above a minimum elevation angle $\theta$, illustrated in Fig.~\ref{fig:system_model} (a). This space is essentially the volume swept by the maximum visible caps, $\mathcal{A}(R(h),h)$, across the entire range of possible satellite altitude $\CMcal{H}$. Its total volume, $|\mathcal{V}|$, can be computed by integrating the area of these caps over the range of heights:
\begin{align}
    |\mathcal{V}| &= \left| \bigcup_{h \in \CMcal{H}} \left\{\mathcal{A}(R(h),h) \right\} \right| \nonumber \\
    &= \int_{\mathcal{V}} r^2 \sin \psi \, d\phi \,d\psi \, dr = \int^{h_{\sf max}}_{0} A(R(h),h) dh.
\end{align}
The formulation uses a spherical coordinate, where $r$ is the radial distance, $\psi$ is the polar angle, and $\phi$ is the azimuth angle.


A significant distinction in visibility between the proposed network model and conventional models \cite{park:twc:22} lies in its joint determination by both spatial location and altitude. Specifically, even if two satellites share the same horizontal position in $\Phi_{\sf S}$, their differing altitudes can lead to one being visible while the other is not, as depicted in Fig.~ \ref{fig:system_model} (a) and (b). Generally, a higher altitude increases the likelihood of a point's visibility from a typical user. The RHM, which we propose, effectively incorporates this crucial characteristic into the system, offering a more simple and practical model for satellite communication systems.

\subsection{Channel Model}
The propagation channel model is composed of both large- and small-scale fading components. The large-scale fading, which accounts for signal attenuation over distance, is modeled as a function of the location of satellites. 
The path loss of a wireless channel between the typical user and the satellite located at $\hat {\mathbf{d}}_i$ is given by
\begin{align}
    \| \hat {\bf{d}}_i - {\bf{u}}_1\|^{-\beta},
\end{align}
where the path loss exponent $\beta$ reflects the characteristics of the wireless environment. For instance, in a channel with a LoS path, $\beta$ is set to 2. 


We model the small-scale fading on each satellite link as an i.i.d. random process following Shadowed-Rician (SR) distribution. This model is well suited for satellite communications as it accurately captures the composite effects of multipath fading and shadowing in both the S and Ka bands \cite{jung:tcom:22, bhatnagar:commlett:2014}. 
The PDF of the fading envelope $\sqrt{X_i}$ for the link from satellite $\hat{\mathbf{d}}_i$ is: 
\begin{align} \label{E:eq:sr}
    & f_{\sqrt{X_i}}(x) =  \!\left(\! \frac{2bm}{2bm + \Omega} \!\right)^m \frac{x}{b}  \!\exp \left(- \frac{x^2}{2b} \right) \! F_1 \left(m; 1; \frac{\Omega x^2}{2b (2bm + \Omega)} \right), 
\end{align}
where $2b$ is the average power of the scattered component, $\Omega$ is the average power of the LoS component, $m$ is the Nakagami parameter, and $F_1(\cdot)$ is the confluent hypergeometric function of the first kind. 
We define $K = {2bm}/{(2bm + \Omega)} $ and $\delta = \Omega/(2b(2bm + \Omega))$. 
For integer $m$, the confluent hypergeometric function is computed as
\begin{align} \label{E:eq:hyper}
    F_1(m; 1; z) = \exp(z) \sum_{k = 0}^{m-1} \frac{(1-m)_k(-z)^k}{(k!)^2},
\end{align}
where $(m)_k$ is the Pochhammer symbol defined as $(m)_k = \Gamma(m+k)/\Gamma(m)$.
Using \eqref{E:eq:hyper}, we obtain the PDF of the Shadowed-Rician fading power as 
\begin{align} \label{E:eq:sr2}
    f_{X_i}(x) =  \left(K \right)^m \!\frac{1}{2b} \! \exp\left(-\frac{x}{2b} \right) F_1 \left(m; 1; \delta x^2 \right). 
\end{align}
From now on, we denote the SR distribution as $\CMcal{SR}(\Omega, b_0, m)$.

We adopt the transmit and receive beamforming gain as the sectored antenna model, wherein the directional beamforming gains are approximated as a rectangular function \cite{renzo:twc:15,bai:commmag:14}. This model approximates the antenna pattern as a rectangular function: a user located within the main lobe experiences a high, constant gain, while a user outside of it receives a low, constant side-lobe gain.
This sectored antenna model has been widely used in stochastic geometry based analysis because it is not only analytically tractable, but it is also suitable to reflect the primary features of the directional beamforming. 
The combined gain product, $G_i$, which incorporates bot transmitter and receiver gains as well as the wavelength-dependent free-space factor, is
\begin{equation}
    G_i = \begin{cases}
    G^{\text{ml}}_{\text{tx}} G^{\text{ml}}_{\text{rx}} \dfrac{c^2}{(4\pi f_c)^2}, & \text{$|\xi_i| < \xi$}, \\
    G^{\text{sl}}_{\text{tx}} G^{\text{sl}}_{\text{rx}} \dfrac{c^2}{(4\pi f_c)^2},  & \text{otherwise}.
    \end{cases}
\end{equation}
Here, $\xi_i$ is the angle between user $i$ and the beam's boresight and $\xi$ is the main-lobe half-beamwidth. $G^{\text{ml}}_{\text{tx}}$ and $G^{\text{sl}}_{\text{tx}}$ are the main-lobe and side-lobe antenna gains at the satellite transmitter, respectively. Similarly, $G^{\text{ml}}_{\text{rx}}$ and $G^{\text{sl}}_{\text{rx}}$ are the main-lobe and side-lobe beamforming gains at the user receiver. For the analysis of a typical user (indexed as $1$), we assume it is located in the main lobe of its serving satellite. Consequently, the desired signal link experiences the main-lobe, which we denote as $G_1$. In contrast, any interfering satellite ($i\neq 1$) is pointing its main lobe elsewhere, placing the typical user in its side-lobe. Thus, the gain for any interfering link is the side-lobe gain, $G_i$.

\subsection{Performance Metric} 
Our primary performance metric is the signal-to-interference-plus-noise ratio (SINR) at a typical user. Without loss of generality, the typical user is served by its closest satellite (indexed as $1$). The SINR of the typical user is characterized by
\begin{align}
    \text{SINR}
    &= \frac{G_1 P_{\sf S}\, X_{1}\, \|\hat{\mathbf d}_{1}-\mathbf u_1\|^{-\beta}}
            {\sum_{\hat{\mathbf d}_i\in \hat{\Phi}_{\sf I}} G_i P_{\sf S}\, X_i\, \|\hat{\mathbf d}_{i}-\mathbf u_1\|^{-\beta} + \kappa T B} \nonumber\\
    &= \frac{X_{1}\, \|\hat{\mathbf d}_{1}-\mathbf u_1\|^{-\beta}}
            {\sum_{\hat{\mathbf d}_i\in \hat{\Phi}_{\sf I}} \bar{G}_i\, X_i\, \|\hat{\mathbf d}_{i}-\mathbf u_1\|^{-\beta} + \sigma^2}, \label{eq:sinr}
\end{align}
where $\kappa$, $T$, and $B$ are the Boltzmann constant, the noise temperature, and the system bandwidth, respectively.
$\beta$ is the path loss exponent. $X_i$ models small-scale fading; 
$\bar{G}_i \triangleq \frac{G_i}{G_1}$ is the interferer-to-serving gain ratio. and $\sigma^2 \triangleq \frac{\kappa T B}{G_1 P_{\sf S}}$.
To simplify subsequent expressions, we define the aggregate interference, $\CMcal{I}$, as:
\begin{align}
    \CMcal{I} \;\triangleq\; \sum_{\hat{\mathbf d}_i\in \hat{\Phi}_{\sf I}} \bar{G}_i\, X_i\, \|\hat{\mathbf d}_{i}-\mathbf u_1\|^{-\beta}.
\end{align}
We evaluated the performance of satellite downlink communication systems by investigating the probability of coverage that the SINR is larger than the threshold $\tau$. 
This probability can be decomposed using the law of total probability as follows:
\begin{align}
     \mathcal{P}_{cov} &= P_{\sf cov}^{\text{SINR}}(\tau, \lambda, f_{H}, R_{\sf S}) = \mathbb{P}[\text{SINR} > \tau] \nonumber\\
     &= \mathbb{P}[\CMcal{N}] \mathbb{P}[\text{SINR} > \tau |\CMcal{N}], \nonumber
\end{align}
where the conditioning event $\CMcal{N}$, is that at least one satellite is visible to the typical user (i.e., $\hat{\Phi}_{\sf S} \cap \mathcal{V} \ne \emptyset$).
This formulation decouples the analysis into two components: the visibility probability, $\CMcal{N}$, and the conditional coverage probability, which assume that the user is not in an outage due to a lack of satellite visibility.

\subsection{Mathematical Preliminaries}

This subsection establishes the key mathematical foundations for our coverage analysis. We first focus on satellite visibility.

\begin{lemma} \label{lem:vis_prob_satellite}
The number of satellites in the visible region $\mathcal{V}$ follows a Poisson distribution. The probability that at least one satellite is visible, $\mathbb{P}[\CMcal{N}]$, is given by
\color{black}{
\begin{align}
     \mathbb{P}[\CMcal{N}] =  1- \exp\left( -\frac{\lambda R_{\sf S}^2}{h_{\max}} \int_{0}^{h_{\max}} \frac{A(R(h),h)}{(R_{\sf S} + h)^2} \, dh \right). \nonumber 
\end{align}}
\end{lemma}
\begin{proof}
See Appendix \ref{proof:lem_vis}. 
\end{proof}
With the visibility probability established, we now characterize the distance to the nearest satellite, which is crucial for determining the desired signal strength.

 \begin{figure}[t!]
    \centering
    \includegraphics[width=1\linewidth]{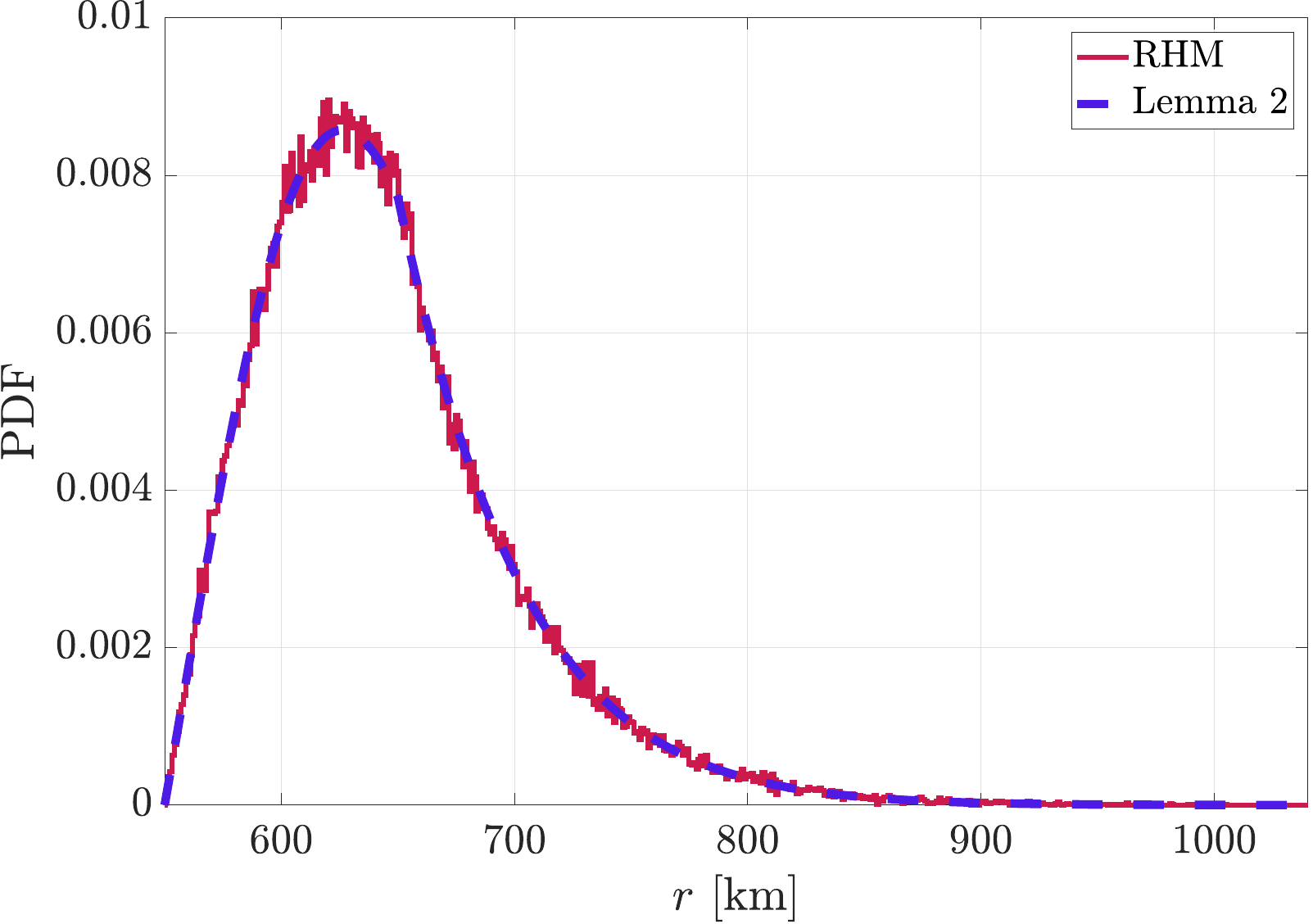}
    \caption{Comparison PDF of the distance to the nearest satellite $R$ where $R_{\sf S} = 550$ km and $h_{\max}$ = 100 km.}
    \label{fig:r pdf}
\end{figure}

\begin{lemma}\label{lem:nearest_satellite}
The PDF of the distance $R$ to the nearest satellite, conditioned on visibility is given by $f_{R| \CMcal{N}}(r)$ as:
\begin{align}
    f_{R| \CMcal{N}}(r) = \frac{\frac{2 \pi r \lambda R_{\sf S}^2}{h_{\max} R_{\sf E} } \int_0^{h(r)} \! \frac{1}{R_{\sf S}+h}\,dh \,  \exp\left(- \frac{\lambda R^2_{\sf S}}{h_{\max}} \int_{0}^{h(r)}\! \frac{{A}(r,h)}{(R_{\sf S} + h)^2} \, dh \right) }{1- \exp\left( -\frac{\lambda R_{\sf S}^2}{h_{\max}} \int_{0}^{h_{\max}} \frac{A(R(h),h)}{(R_{\sf S} + h)^2} \, dh \right)}.
\nonumber 
\end{align} 
\end{lemma}
 \proof 
See Appendix \ref{proof:lem_nearest}. 
\endproof
The integral limit $h(r)$ in the expression is defined as $h(r) = \min(r- h_{\sf S},h_{\max})$. 

Fig.~\ref{fig:r pdf} plots the PDF of the nearest satellite distance $R$, comparing our analytical expression from Lemma \ref{lem:nearest_satellite} (dotted line) with results from Monte Carlo simulations (dashed line). The figure provides clear validation of our model, as our analysis perfectly matches the simulation results.
The shape of the distribution is physically intuitive. Although satellites are uniformly distributed in a spherical shell concentric with the Earth's origin, the user is on the surface and therefore offset from this center. This asymmetrical viewpoint results in a skewed probability distribution for the distance R, concentrating the likelihood in a specific range. Furthermore, this generalized distribution is consistent with prior work; it resembles a truncated Rayleigh distribution and correctly reduces to the known result for a single-altitude model \cite{park:twc:22} as $h \to 0$.


Now we derive the coverage probability of the RHM based on the preliminaries from the previous section. A key step in this derivation is to characterize the aggregate interference. We achieve this by finding its conditional Laplace transform, which is presented in the following lemma.
\begin{lemma}\label{lem:int_laplace}
The conditional Laplace transform of the aggregate interference $\CMcal{I}$, given that all interfering satellites are at a distance greater than r from the typical user, is
\begin{align}
   &\CMcal{L}_{\CMcal{I}_r|\CMcal{N}}(s) = \exp \left( - \frac{2\pi\lambda R^2_{\sf S}}{h_{\max} R_{\sf E}} \int_0^{h_{\sf max}} \frac{1}{R_{\sf S}+h}  \right. \nonumber \\
    &\quad \times \int_{\max(r,h_{\sf S}+h)}^{R(h)} \left. \left(1 - \frac{1}{(1 + s \eta \bar{G}  v^{-\beta})^{\alpha}} \right) v \, dv \, dh \right). \nonumber
\end{align}
\end{lemma}
\proof 
See Appendix \ref{proof:laplace}. 
\endproof
The double integral captures the total interference by integrating over all distances $v$ at a fixed height $h$, and then over all possible heights. 
For any fixed $h$, all potentially visible satellites reside on the spherical cap $\mathcal{A}(R(h),h)$. However, by definition, an interfering satellite must be farther away than the serving satellite, which is at distance $r$.
 In other words, the integration is performed only over the portion of the visible cap that lies outside the exclusion radius $r$ around the user.
In this regard, the lower limit, $\max(r,h_{\sf S}+h)$, correctly defines the region of interferers. Since the serving satellites is at distance $r$, all interferers must be farther away. Simultaneously, for any satellite at height $h$, the minimum possible LoS distance is $h_{\sf S}+h$. Therefore, the integration for interference at a given height must start from the larger of these two minimums, effectively excluding the serving link and respecting the systems' geometry, as depicted by Fig.~\ref{fig:system_model} (d).
 


\section{Rate Coverage Analysis}
In this section, we derive the rate coverage probability for the RHM. Our approach involves two key steps: first, we formulate an exact expression based on a Gamma approximation of the fading channel, and second, we derive a more tractable form using a tight analytical bound.


\subsection{Gamma Approximation}

The inherent complexity of the SR fading model makes a direct derivation of the coverage probability intractability. Therefore, following a widely adopted methodology for satellite communication analysis, we approximate the SR fading power with a Gamma distribution \cite{talga:taes:2024, abdi:twc:03}. This approach is not only analytically tractable but also highly accurate, as the Gamma distribution's parameters are matched to the moments of the SR distribution, precisely capturing the channel's essential characteristic.

\begin{proposition}\label{prop:gamma dist}
    Assuming that the SR fading follows $\sqrt{X} \sim \CMcal{SR}(\Omega, b_0, m)$, the power envelope $X$ can be accurately approximated by a Gamma distribution $\Gamma(\alpha, \eta)$, whose PDF is given by
    \begin{align}
    f_{X}(x) \approx \frac{1}{\Gamma(\alpha), \eta^\alpha} x^{\alpha - 1} \exp\left(- \frac{x}{\eta} \right),
    \end{align}
    where the shape and scale parameters are defined as $\alpha = \frac{m (2b + \Omega)^2}{4mb^2 + 4mb\Omega + \Omega^2}$ and $\eta = \frac{4mb^2 + 4mb\Omega + \Omega^2}{m(2b + \Omega)}$, respectively.  
\end{proposition}

\subsection{Coverage Probability}
Using Proposition \ref{prop:gamma dist}, we derive an exact expression for the coverage probability.
\begin{theorem} \label{thm:coverage prob}
The rate coverage probability of RHM is represented as
\begin{align}
    &\mathcal{P}_{cov} = P_{\sf cov}^{\text{SINR}}(\tau, \lambda, f_{H}, R_{\sf S}) \nonumber \\
    &= \mathbb{P}[\CMcal{N}] \int_{h_{\sf S}}^{R(h_{\sf max})} f_{R|\CMcal{N}}(r) \sum_{k = 0}^{\alpha - 1} \frac{(-s)^k}{k!} \cdot \frac{\partial^k \CMcal{L}_U(s)}{\partial s^k} \Bigg|_{s = \frac{r^\beta \tau}{\eta}} dr, \nonumber
\end{align}
where $\CMcal{L}_{U_r}(s) = e^{-s\sigma^2} \CMcal{L}_{\CMcal{I}_r}(s)$. $\mathbb{P}[\CMcal{N}]$ and $f_{R\mid\CMcal{N}}(r)$ are given in Lemmas~\ref{lem:vis_prob_satellite} and~\ref{lem:nearest_satellite}, respectively.
\end{theorem} 
\proof
See Appendix \ref{proof:thm:coveage prob}. 
\endproof

\begin{table}[t!]
\centering
\caption{Satellite Parameters} 
\label{tab:simul_para}
\begin{tabular}{|>{\Centering\arraybackslash}m{1.6cm}|>{\Centering\arraybackslash}m{1.2cm}|>{\Centering\arraybackslash}m{1.2cm}|>{\Centering\arraybackslash}m{1.4cm}|>{\Centering\arraybackslash}m{1.4cm}|}
\hline
\multirow{2}{*}{\makecell{\textbf{Parameters}}} & \multicolumn{4}{c|}{\textbf{Satellites}} \\ \cline{2-5}
& \text{Starlink} & \text{OneWeb} & \text{Globalstar} & \text{Strela} \\ \hline \hline
$R_{\sf S}$ & 550 km & 1200 km & 1400 km & 1400 km \\ \hline
$\beta$ & \multicolumn{4}{c|}{2} \\ \hline
$G^{\text ml}_{\text tx}$ & 44 dBi & 37 dBi & 20 dBi & 20 dBi \\ \hline
$G^{\text sl}_{\text rx}$ & 24 dBi & 20 dBi & 0 dBi & 0 dBi \\ \hline
$\kappa$ & \multicolumn{4}{c|}{-228.6 dBW/K/Hz} \\ \hline
$T$ & \multicolumn{4}{c|}{120 K} \\ \hline
$B_w$ & 250 MHz & 250 MHz & 16.5 MHz & 16.5 MHz \\ \hline
Frequency Band & Ku/Ka & Ku/Ka & S & S \\ \hline
$f_c$ & 12 GHz & 12 GHz & 2500 MHz & 2500 MHz \\ \hline
$P$ & 40 dBm & 35 dBm & 35 dBm & 35 dBm \\ \hline
$N$ & 8245 & 648 & 85 & 565 \\ \hline
Visible Minimum Angle & 25\degree & 15\degree & 10\degree & 10\degree \\ \hline
\end{tabular}
\end{table}


The exact expression in Theorem \ref{thm:coverage prob}, while analytically precise, is computationally demanding due to high-order derivatives. Therefore, to derive a more tractable result, we adopt a methodology that has proven effective in the literature: approximating the distribution using a tight bound on its CCDF \cite{talga:taes:2024, abdi:twc:03}. This allows us to develop a new, compact expression for the coverage probability that avoids derivatives entirely, making highly suitable for numerical analysis.
\begin{proposition}\label{prop:tight alzer bound}
    For given Gamma RV $X \sim \Gamma(\alpha, \eta)$, the CCDF of the Gamma distribution is tightly bounded as 
    \begin{align}
        &F^c_X(x) \ge 1- (1-e^{-\frac{\mu x}{\eta}})^{\alpha}, \;\;\; \text{if $\alpha \le 1$}, \\
        &F^c_X(x) \le 1- (1 - e^{-\frac{\mu x}{\eta}})^{\alpha}, \;\;\; \text{if $\alpha > 1$},
    \end{align}
    where $\mu = (\alpha !)^{-\frac{1}{\alpha}}$.
\end{proposition}
The bounds presented in Proposition~\ref{prop:tight alzer bound} are a direct consequence of the well-known Alzer's inequality, which bounds the incomplete Gamma function \cite{alzer:mathcomp:1997}:
\begin{align}
    (1-e^{-\mu \alpha x })^{\mu} \le \frac{1}{\Gamma(\mu)} \int^{\mu x}_0 t^{\mu - 1} e^{-t} dt \le (1-e^{-\mu x})^{\mu}
\end{align}
where $\mu = (\alpha !)^{-\frac{1}{\alpha}}$. Using this, the CCDF of gamma distribution is tightly bounded as in Proposition \ref{prop:tight alzer bound}. 
With this, we compute the coverage probability in a simple form.
\begin{corollary}\label{coro: p cov gam approx}
    Under the tight bound in Proposition \ref{prop:tight alzer bound}, the rate coverage probability of RHM is given by
    \begin{align}
        &\bar{\mathcal{P}}_{cov} = \bar P_{\sf cov}^{\text{SINR}}(\tau, \lambda, f_{H}, R_{\sf S}) \nonumber \\
        &= \mathbb{P}[\CMcal{N}] \int^{R(h_{\max})}_{h_{\sf S}} f_{R|\CMcal{N}}(r) \sum_{k=1}^{ \alpha} \binom{ \alpha}{k}(-1)^{k+1} \CMcal{L}_{U_r}\left(\frac{k\mu r^\beta \tau}{\eta}\right)\,dr. \nonumber
    \end{align}
\end{corollary}
\begin{proof}
    Please see Appendix \ref{proof: p cov gam approx}
\end{proof}
Crucially, the result in Corollary~\ref{coro: p cov gam approx} replaces the challenging derivative operations with a simple weighted sum of Laplace transform evaluations. This provides computationally efficient, and highly accurate method for performance analysis, which we will verify in the subsequent section.


\begin{figure}[t!]
    \centering
    \includegraphics[width=1\linewidth]{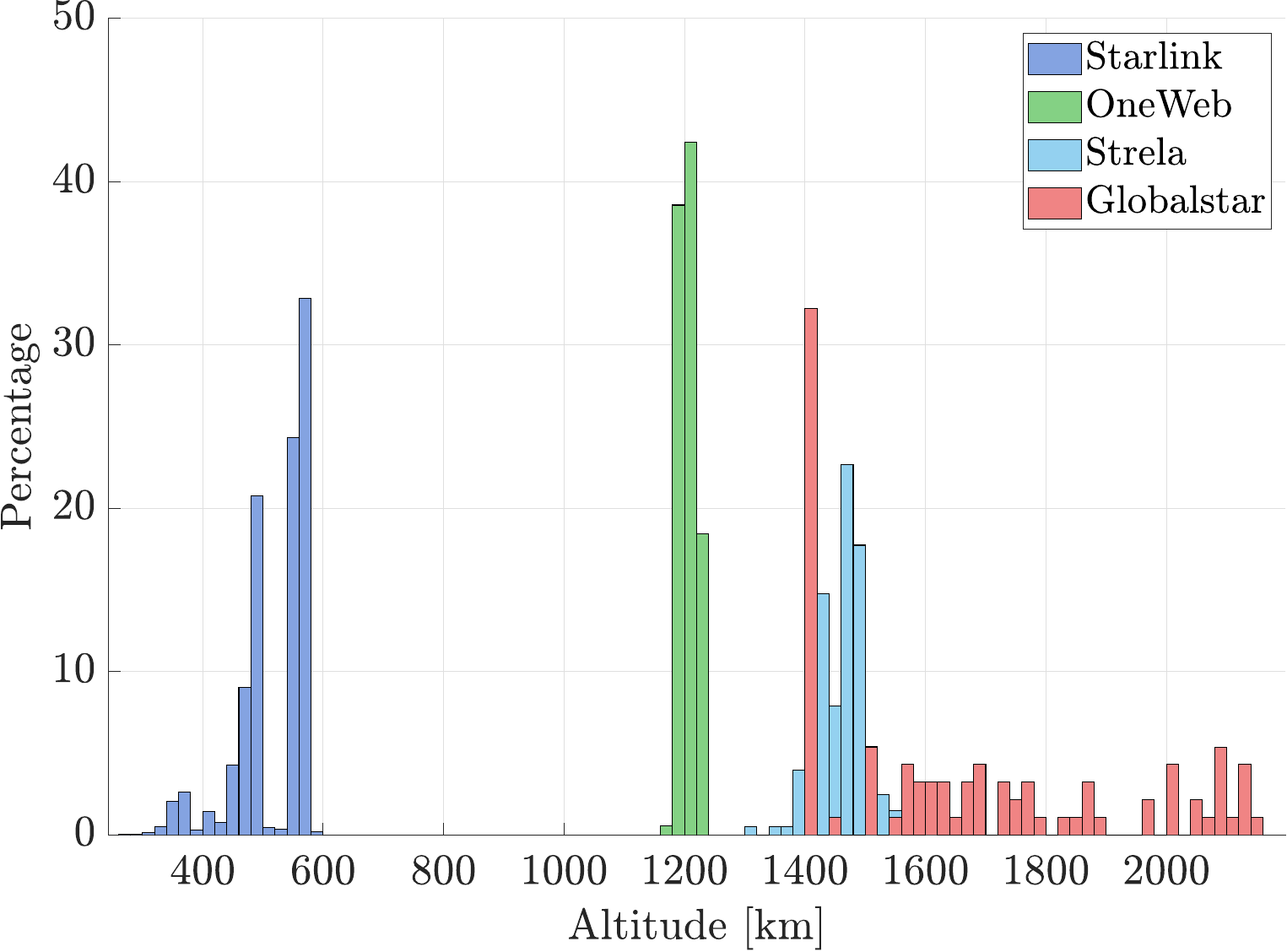}
    \caption{Histogram of altitude of commercial satellite in operation.}
    \label{fig:satellite altitude}
\end{figure}
\begin{figure}[t!]
    \centering
    \includegraphics[width=1\linewidth]{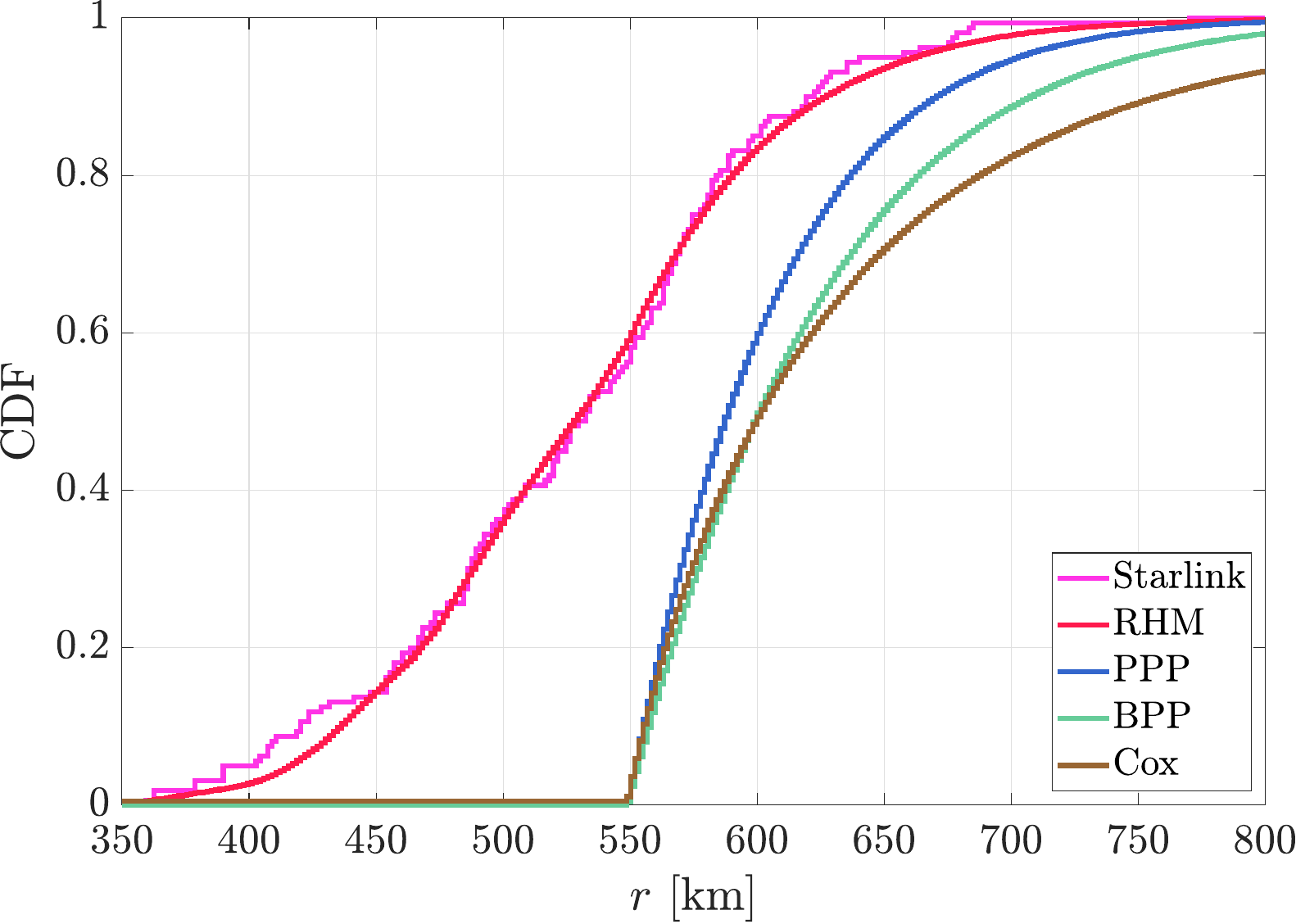}
    \caption{Comparison of CDFs for the nearest satellite distance $r$ between the Starlink constellation and single-altitude models at $R_{\sf S} = 550$ km.}
    \label{fig:r cdf}
\end{figure}

\section{Numerical Results} \label{section: numerical results}

In this section, we demonstrate the suitability of the proposed RHM for modeling real-world satellite constellations by comparing its coverage expressions with those of commercial LEO systems. We also validate our analytical expressions and approximation accuracy through Monte Carlo simulations. The system parameters adopted in our analysis are summarized in Table~\ref{tab:simul_para}, and are chosen to be consistent with recent works \cite{Hui:iotj:2025, okati:tcom:20, park:twc:22}.

\begin{figure*}[!t]
  \centering
  \subfloat[]{%
    \includegraphics[width=0.48\textwidth]{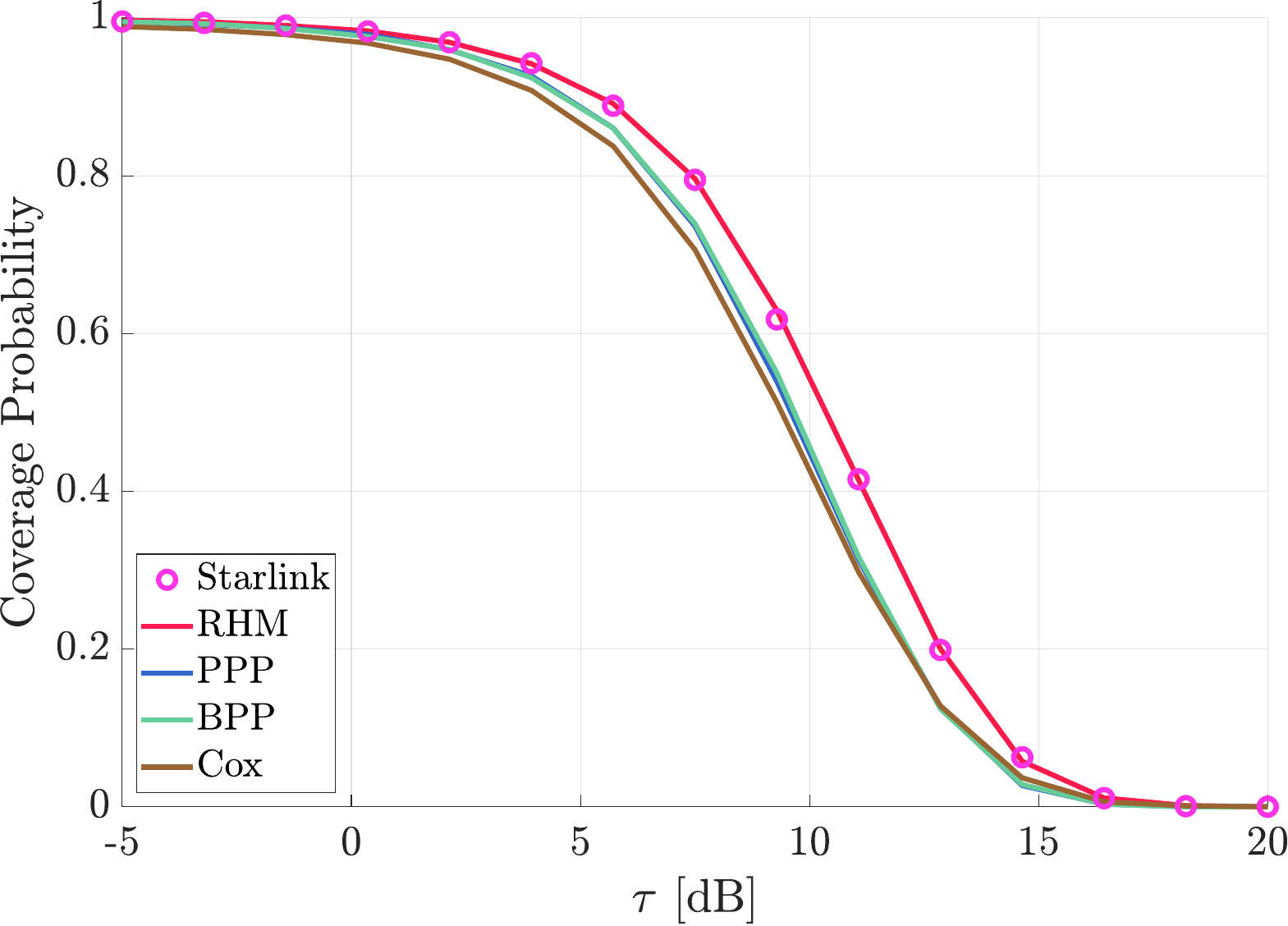}%
    \label{fig:starlink}%
  }
  \hfill
  \subfloat[]{%
    \includegraphics[width=0.48\textwidth]{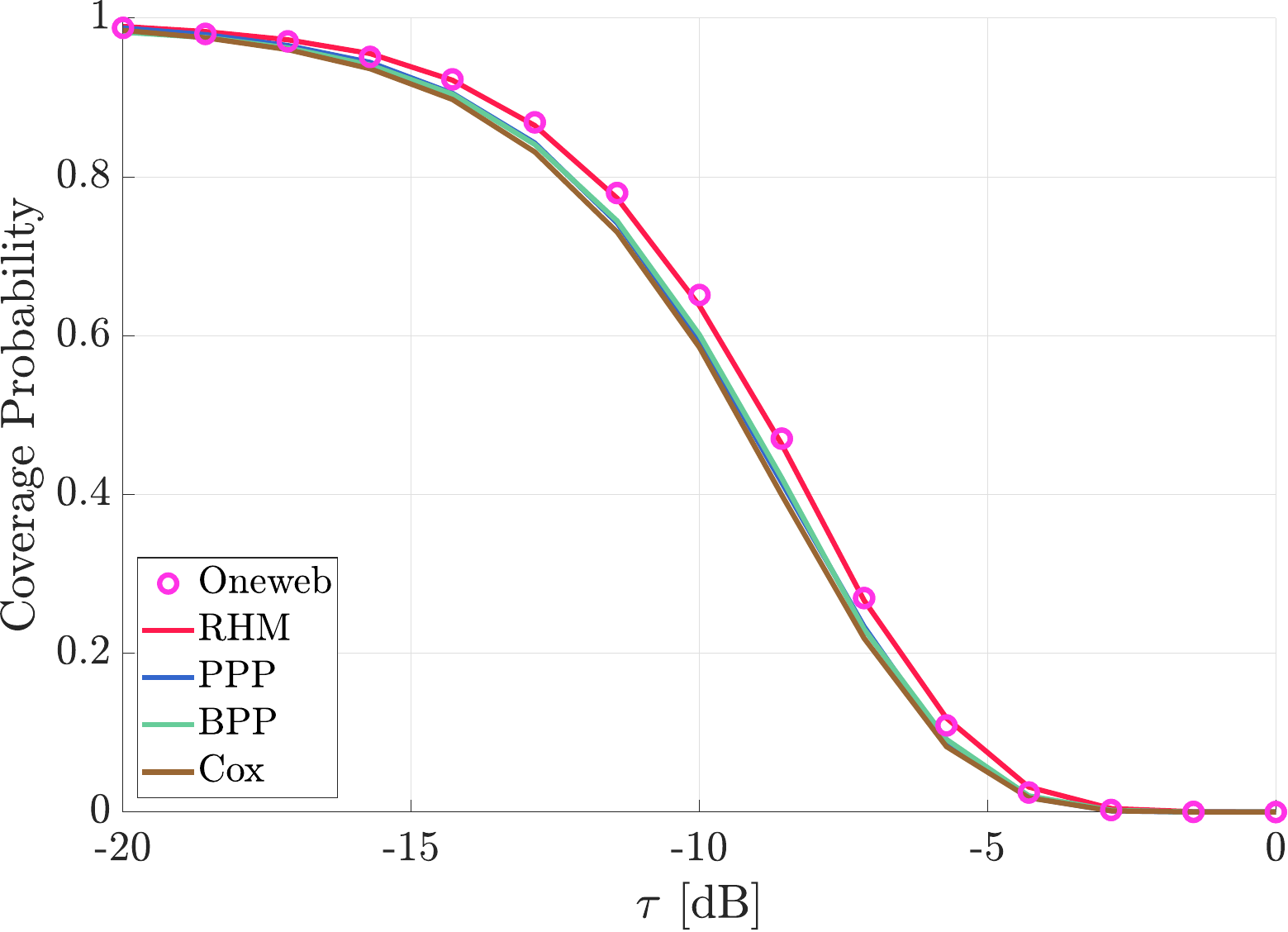}%
    \label{fig:OneWeb}%
  }
  \caption{Coverage probability versus SINR threshold $\tau$ for (a) Starlink and (b) OneWeb, compared with the proposed RHM and conventional single-altitude models (PPP, BPP, and Cox).}
  \label{fig:satellite coverage starlink OneWeb}
\end{figure*}

\begin{figure*}[!t]
  \centering
  \subfloat[]{%
    \includegraphics[width=0.48\textwidth]{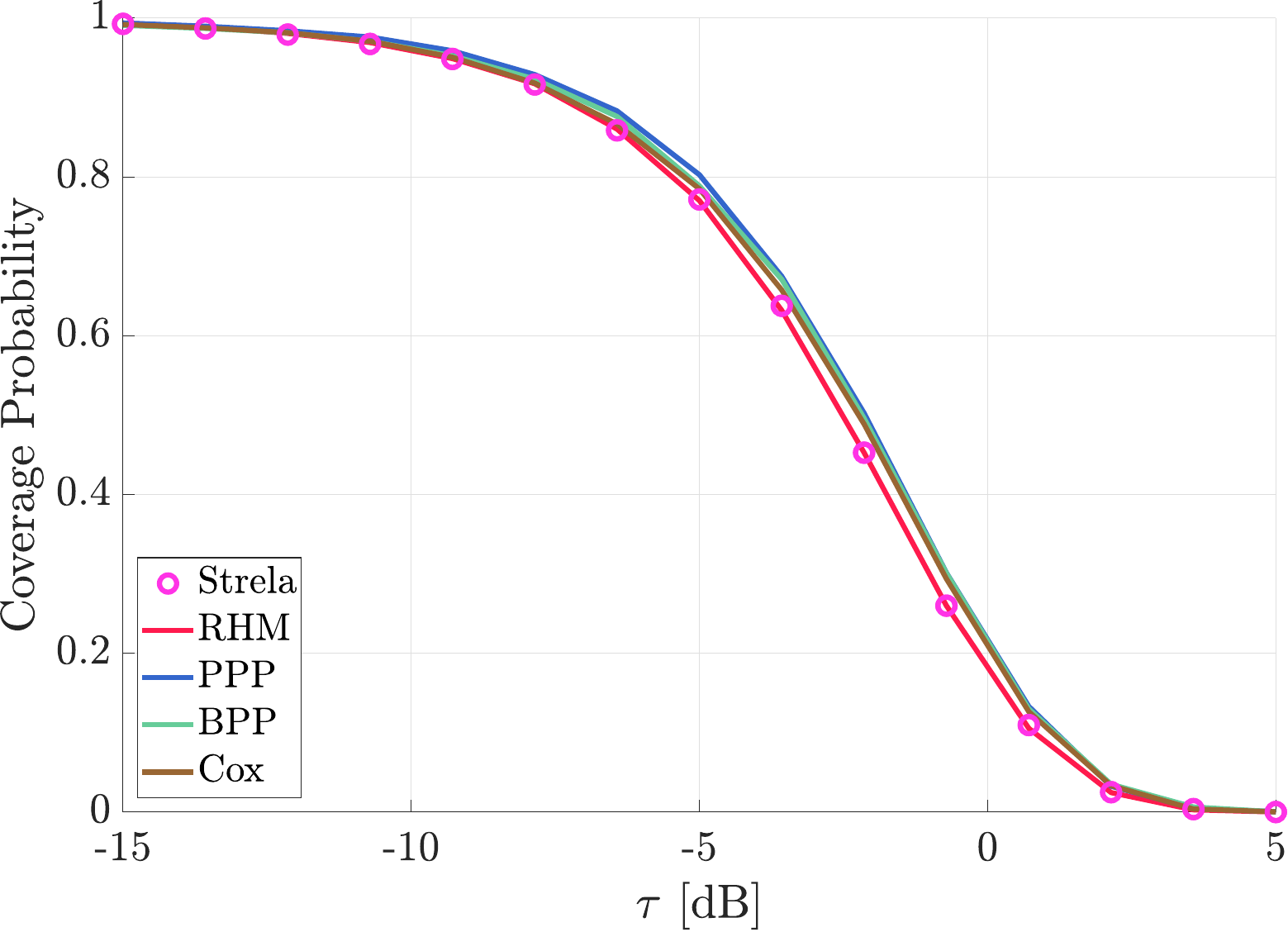}%
    \label{fig:strela}%
  }
  \hfill
  \subfloat[]{%
    \includegraphics[width=0.48\textwidth]{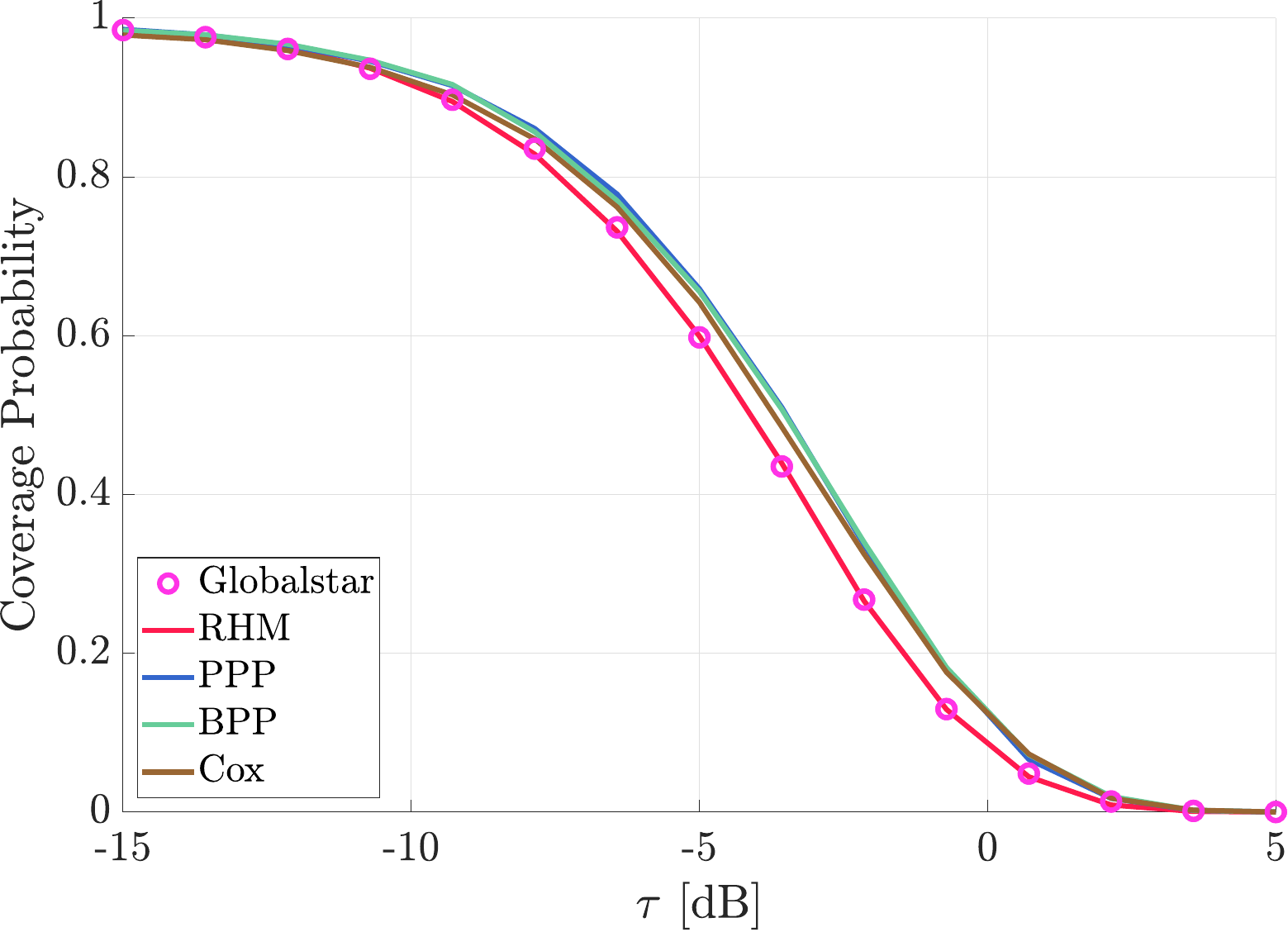}%
    \label{fig:globalstar}%
  }
  \caption{Coverage probability versus SINR threshold $\tau$ for (a) Strela and (b) Globalstar, compared with the proposed RHM and conventional single-altitude models (PPP, BPP, and Cox).}
  \label{fig:satellite coverage strela globalstar}
\end{figure*}

\subsection{Satellite Altitude Distribution}
We highlight the necessity of the RHM by examining the altitude distributions of operational satellite constellations from the website: https://www.n2yo.com. 
The foundational assumption in many satellite network analyses is that the satellite orbits at a single-altitude \cite{talgat:commlett:20, okati:tcom:20, okati:pimrc31:2020, homssi:commlett:2021, park:twc:22, lee:access:2024}. However, data from operational constellations reveal a more complex reality. 

Fig.~\ref{fig:satellite altitude} shows the histogram of altitude for the commercial satellite operators, where constellations exhibit varied altitude distributions. For instance, Starlink has a wide distribution form 350 km to 580 km, while OneWeb's is highly concentrated around 1200 km.
This initial comparison highlights that constellations differ significantly in both their mean altitude and the variance of their distribution. To more directly isolate the impact of the distribution shape, it is particular insightful to compare constellations that share a similar mean altitude but differ in their dispersion.

An even more compelling case is the comparison between Strela and Globalstar. This pair is particularly insightful because both constellations share a similar mean altitude of approximately 1400 km, allowing us to isolate the impact of the distribution's shape. As the figure shows, their distributions are markedly different: The Strela constellation is tightly clustered, with most satellites in a narrow 1300–1500 km band. 
In contrast, the Globalstar constellation is broadly dispersed, with a significant number of satellites spread across a wide range extending to 2200 km. 
In the next subsection, we demonstrate how this distributional difference critically impacts performance and how the RHM accurately captures this effect.

In Fig.~\ref{fig:r cdf}, we compare the empirical CDF of the closest satellite distance for Starlink with several analytical models.
For this comparison, our RHM incorporates Starlink's actual altitude distribution, while the single-altitude models (PPP, BPP, and Cox) assume a nominal altitude of $R_{\sf S} = 550$ km. 
The RHM and PPP models are parameterized by the average satellite intensity $\lambda$, whereas the BPP and Cox models are initially parameterized using the actual number of deployed satellites, $N$, as listed in Table~\ref{tab:simul_para}.
A key observation is that the CDFs for the BPP and Cox models deviate significantly from the empirical Starlink data, showing a much larger discrepancy than the PPP-based models. 
Therefore, to ensure a fairer comparison based on equivalent satellite density in our subsequent performance analysis, we normalize the BPP and Cox models by setting their satellite count according to the relation $N = 4\pi R^2_{\sf S}\lambda$. 
While a similar CDF trend does not guarantee an identical PDF, the CDF provides a suitable basis for comparison, particularly given the discrete nature of orbital shells in real-world constellations.


\begin{table}[t!]
{\color{black}{
  \begin{center}
    \caption{Shadowed-Rician Channel Parameters}
    \label{tab:shadowing} 
    \begin{tabular}{c|c|c|c} 
    \hline
    {\textbf{Scenarios}} & {$m$} & {$b$} & {$\Omega$}  \\
    \hline \hline
      {Frequent Heavy Shadowing (FHS)} & {$1$} & {$0.063$} & {$8.97\times 10^{-4}$}   \\
    \hline
      {Average Shadowing (AS)} & {$10$} & {$0.126$} & {$0.835$} \\
    \hline
    {Infrequent Light Shadowing (ILS)} & {$19$} & {$0.158$} & {$1.29$} \\
   \hline 
    \end{tabular}
  \end{center}
  }}
\end{table}
\subsection{RHM Validation}

In this subsection, we show how our PPP-based RHM is a suitable analytical tool for real-world satellite communication systems by comparing its coverage probability with that of commercial satellite constellations. We also demonstrate the validity of our analysis and approximations through Monte Carlo simulations. The system parameters used in our simulations are summarized in Table~\ref{tab:simul_para}, aligned with the values provided in \cite{Hui:iotj:2025, okati:tcom:20, park:twc:22}.

Fig.~\ref{fig:satellite coverage starlink OneWeb} (a) and (b) illustrate the coverage probabilities of the actual Starlink and OneWeb constellations, along with those simulations of stochastic geometry-based models, including our proposed RHM. As the figure shows, both Starlink and OneWeb exhibit a significant performance gap when compared to the conventional single-altitude stochastic models, PPP, BPP and Cox. Especially, for Starlink at threshold $\tau$ of approximately 11 dB, the PPP shows a maximum error rate of 24\%, the BPP shows a maximum error rate of 23.6\%, and the Cox model shows a 26\% error rate.
In contrast, our RHM achieves a minimal error rate of just 0.9\%. 
Similarly, for OneWeb at a $\tau$ of around -8.6 dB, the PPP, BPP and Cox models show error rates of 10.7\%, 10.5\% and 16.8\%, respectively, while the RHM demonstrates a mere 0.6\% error rate. These results validate that our proposed RHM accurately models the performance of real-world satellites that are distributed across various altitudes.
A notable observation here is that the performance discrepancy with single-altitude models is more evident for Starlink than for OneWeb. To understand this difference, we refer back to Fig.~\ref{fig:satellite altitude}, which shows that Starlink's altitude distribution is significantly more dispersed than OneWeb's.

The impact of satellite altitude distribution on overall performance is even more evident in Fig.~\ref{fig:satellite coverage strela globalstar} (a) and (b), which show the coverage probabilities for the Strela and Globalstar constellations as a function of the threshold $\tau$. Both constellations have a similar mean altitude of approximately 1400 km, but their distributions are markedly different. The Strela constellation is tightly clustered around 1400 km, whereas Globalstar has a concentration of satellites at a similar altitude but also a broad distribution of satellites ranging from 1400 km to 2200 km.

This difference in distribution results in a clear performance discrepancy. For the tightly clustered Strela constellation, the performance gap between the actual data and the single-altitude PPP and Cox models is relatively small. Specifically, at $\tau$ of $-3.5$ dB, the error rates with the PPP, BPP, and Cox model are 5.8\%, 5.2\%, and 3.1\%, respectively. In contrast, the error rates for Globalstar with the PPP, BPP, and Cox model at the same $\tau$ enhance  17\%, 16.5\%, and 11.4\%, respectively. 
However, our RHM is able to close this gap, achieving an impressively low error rate of just 0.9\% for Strela and Globalstar at the same threshold. It is worth noting that due to various real-time factors (e.g., collision avoidance, seamless coverage), the positions of actual satellites are constantly and optimally adjusted, making it impossible for any static stochastic model to perfectly track their performance. Although our RHM cannot replicate the actual coverage probability perfectly, it clearly demonstrates that satellite altitude distribution has a significant impact on performance, and our model is capable of accounting for it, with altitude distribution.



\begin{figure}[t!]
    \centering
    \includegraphics[width=\linewidth]{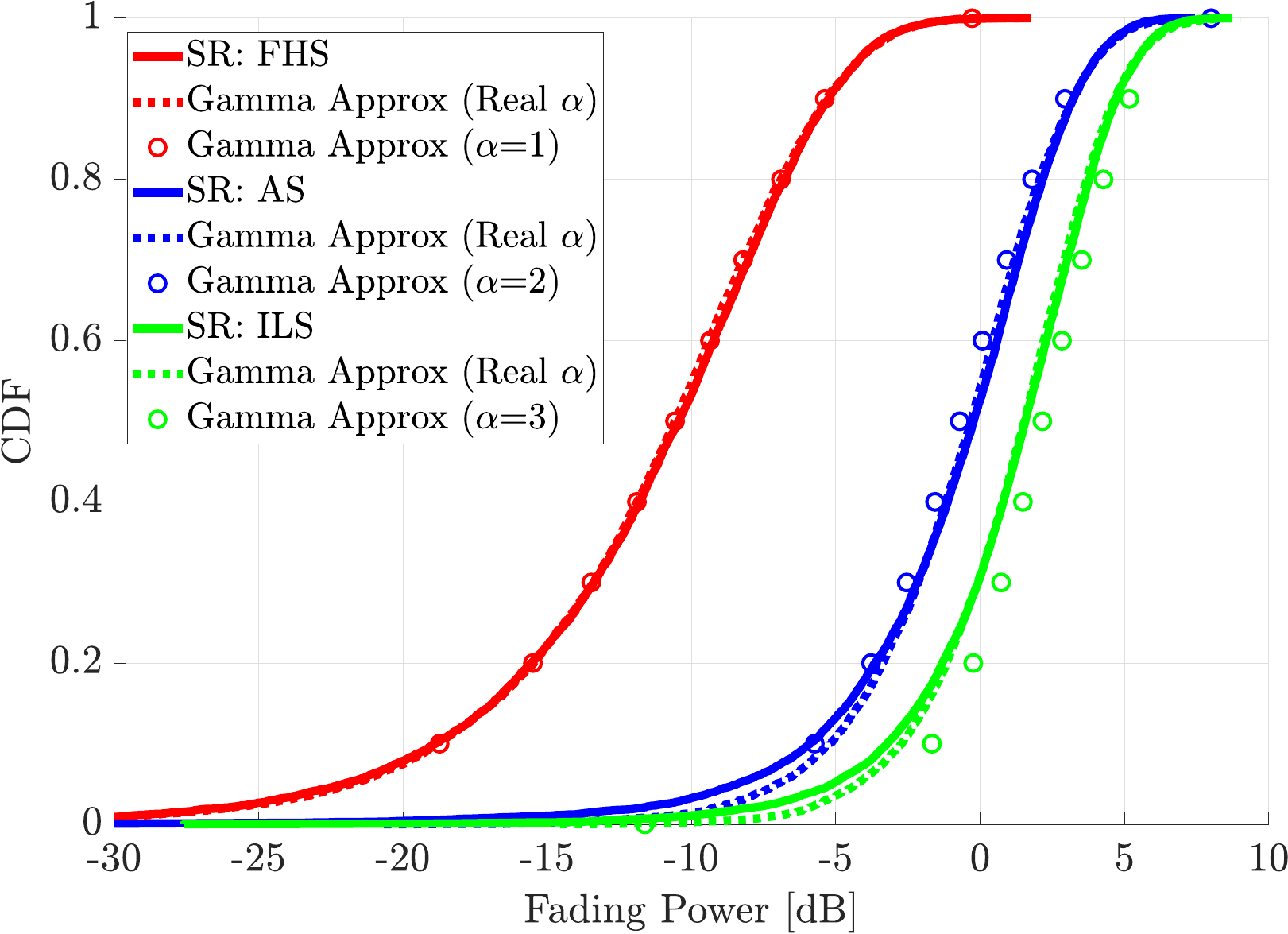}
    \caption{Comparison of Shadowed-Rician fading and Gamma approximation as integer $\alpha$.}
    \label{fig:gamma}
\end{figure}

\begin{figure}[t!]
    \centering
    \includegraphics[width=1\linewidth]{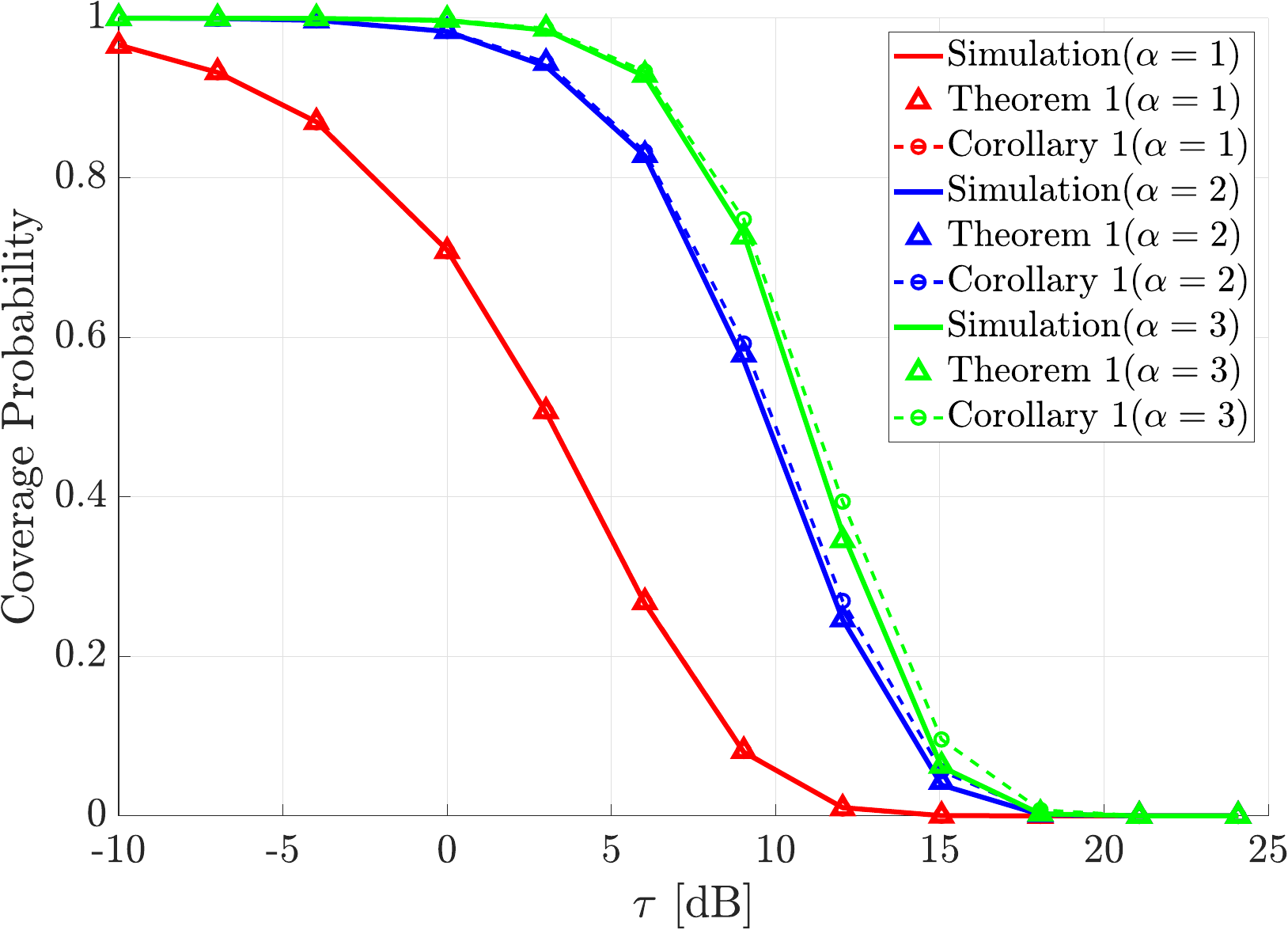}
    \caption{Coverage probability versus threshold $\tau$ comparing simulation with analysis and approximation for different $\alpha$.}
    \label{fig:cov prob}
\end{figure}

\subsection{Accuracy of Analytical Expressions} \label{subsection:accuracy}

In this subsection, we verify the accuracy of our main analytical expression: the exact coverage probability in Theorem \ref{thm:coverage prob} and its tractable approximation in Corollary \ref{coro: p cov gam approx}. This validation proceeds in two logical steps. We first justify the underlying Gamma approximation for the fading channel, and then we compare our final analytical results against Monte Carlo simulations of the RHM.

The first step is to validate the Gamma approximation for the SR fading channel. As shown in Fig.~\ref{fig:gamma}, the Gamma approximation from Proposition \ref{prop:gamma dist} provides a tight fit to the SR distribution, which is why a useful approximation is widely adopted in the literature \cite{jia:iotj:22, talga:taes:2024, abdi:twc:03}. For our analysis, we consider three distinct SR channel scenario, corresponding to the parameters given in Table~\ref{tab:shadowing}. For a tractable derivation, we follow the approach of \cite{talga:taes:2024} and round the shape parameter $\alpha$ to the nearest integer. Fig.~\ref{fig:gamma} confirms that even with this integer approximation, the Gamma CDF still closely tracks the original SR distribution, validating it user in our framework.

With the accuracy of the Gamma approximation established, Fig.~\ref{fig:cov prob} validates our final analytical results against Monte Carlo simulations of the RHM, where we set $R_{\sf S} = 550$ km and $h_{\sf max}=100$ km. The results confirm the high fidelity of our analytical framework. 
First, we confirm the accuracy of our exact expression from Theorem~\ref{thm:coverage prob}. 
Our rigorous analysis from Theorem~\ref{thm:coverage prob} exhibits a negligible difference from the simulation, exhibiting a negligible difference and thus validating the correctness of our formal derivation.
Building on this, our simplified, low-complexity expression from Corollary~\ref{coro: p cov gam approx} also demonstrates remarkable accuracy. While it introduces a marginal gap as an inherent trade-off for its computational simplicity, the error remains minimal across all scenarios.
For instance, at $\tau = 3$ dB, the error is merely 1.9\% for $\alpha=1$, 3.7\% for $\alpha =2$, and 2.5\% for $\alpha=3$.
The slight increase in this error as $\alpha$ grows is a natural consequence of approximating the $\alpha$-order derivative from the exact theorem. This confirms that our corollary is not only a practical but also a highly accurate tool for performance evaluation, successfully validating both our rigorous and simplified approaches.

This analytical validation, combined with the empirical evidence, confirms our RHM framework is a novel, accurate, and reliable tool for analyzing the performance of modern, multi-altitude satellite networks.

\section{Conclusions}
In this paper, we presented PPP-based stochastic geometry to analyze the downlink coverage probability of 3D constellations. We introduced a novel Random Height Model by considering a Poisson point process on a sphere and applying a random radial height to each point. Based on this model and incorporating a sophisticated Shadowed-Rician fading channel, we successfully characterized fundamental network properties, such as visibility probability, nearest satellite distance, and the Laplace transform of the interference. This enabled the derivation of an exact downlink coverage probability, for which we also provided a simplified and tractable expression by applying Alzer’s inequality. The practical relevance and accuracy of our RHM were demonstrated by showing its close alignment with the performance of real-world commercial constellations.
Finally, the accuracy of our analytical expressions was rigorously validated against Monte Carlo simulation results.

\appendices
\section{Proof of Lemma \ref{lem:vis_prob_satellite}} \label{proof:lem_vis}
For a Poisson point process $\Phi_{\sf S}$ with intensity $\lambda$, we introduce a marked process $\hat{\Phi}^{h}_{\sf S}$ as the subset of $\hat{\Phi}_{\sf S}$ containing all points with height $h$.
Then, $\hat{\Phi}^{h}_{\sf S}$ is a homogeneous PPP with density of
\begin{align}\label{eq:lambda}
    \Delta\lambda (h) = \lambda \frac{R_{\sf S}^2}{(R_{\sf S} + h)^2} f_{H}(h) \Delta h.
\end{align}
Based on this, a satellite at $\hat {\bf{d}}_i$ (with height $h$) is visible to the typical user if and only if $\hat {\bf{d}}_i \in \mathcal{A}(R(h),h)$. 
Accordingly, the probability that no satellite with height $h$ exists in $\mathcal{A}(R(h),h)$ is given by
\begin{align}\label{eq:prob thin}
    \mathbb{P}[\hat \Phi_{\sf S}^h (\mathcal{A} (R(h),h))=0] = \exp \left(  -\Delta \lambda(h) |\mathcal{A}(R(h),h)| \right). 
\end{align}
Based on this, the void probability is given by
\begin{align}
    \mathbb{P}[\hat \Phi_{\sf S}=0] &=\mathbb{P}\left[\bigcap_{h \in \CMcal{H}_S} \hat \Phi_{\sf S}^h (\mathcal{A}(R(h),h) )=0 \right] \nonumber \\  
    &\mathop{=}^{\text{(a)}} \prod_{h\in \CMcal {H}_s} \exp \left( -\Delta \lambda(h) A(R(h),h) \right) \nonumber \\
    &\mathop{=}^{\text{(b)}}\exp\left( - \frac{\lambda R_{\sf S}^2}{h_{\max}} \sum_{h\in\CMcal {H}} \frac{A(R(h),h)}{(R_{\sf S} + h)^2} \, \Delta h \right) \nonumber \\
    &\mathop{=}^{\text{(c)}}\exp\left( -\frac{\lambda R_{\sf S}^2}{h_{\max}} \int_{h\in\CMcal {H}} \frac{A(R(h),h)}{(R_{\sf S} + h)^2} \, dh \right), \nonumber
\end{align}
where (a) holds from \eqref{eq:prob thin} because $\hat{\Phi}^{h_1}_{\sf S}$ and $\hat{\Phi}^{h_2}_{\sf S}$ are statistically independent if $h_1 \neq h_2$ due to independent thinning. (b) is from \eqref{eq:lambda} and $f_H(h) = \frac{1}{h_{\sf max}}$. (c) comes from Riemann integration with $\Delta h \to 0$.
Hence, $\CMcal N$ represents there is at least one satellite visible to typical user, we have
\begin{align}
    \mathbb{P}[\CMcal{N}] = 1- \exp\left( -\frac{\lambda R_{\sf S}^2}{h_{\max}} \int_{h\in\CMcal {H}} \frac{A(R(h),h)}{(R_{\sf S} + h)^2} \, dh \right),
\end{align}
where $R(h)$ and $A(R(h),h)$ are as defined in \eqref{eq:R(h)} and \eqref{eq:A(R,h)}, respectively
This completes the proof.

\section{Proof of Lemma \ref{lem:nearest_satellite}} \label{proof:lem_nearest}

Denoting $R$ by the nearest satellite distance to the typical user, the event $R > r$ is equivalent to the event that there is no satellite with a height $h$ in $\mathcal{A}(R(h),h)$. 
Denoting $\hat{\Phi}_{\sf S}^h$ as a set of satellites with height $h$, we represent the CCDF of $R$ conditioned on the fact that at least one satellite is visible to the typical user as 
\begin{align}
    &F_{R | \CMcal{N} }^{\text c}(r) =  \mathbb{P} [R > r |  \CMcal{N}] \nonumber \\
    & \mathop{=}^{(a)} 
    \frac{ \left( \begin{array}{l} {\mathbb{P}[ \bigcap_{h \in \CMcal{H}} \left\{ \hat \Phi_{\sf S}^{h}(\mathcal{A}(r,h)) = 0\right\}] \cdot} \\ 
    \left\{(1- \mathbb{P}[ \bigcap_{h \in \CMcal{H}} \hat \Phi_{\sf S}^{h}(\mathcal{A}(r,h) \backslash \mathcal{A}(r,h)) =0 ])\right\} \end{array} \right) }{\mathbb{P}[ \CMcal{N}]}, \label{eq:nearest_computable}
\end{align}
where (a) follows from the fact that the PPP in $\mathcal{A}(r,h)$ and $\mathcal{A}(R(h),h) \backslash \mathcal{A}(r,h)$ are independent since their sets do not overlap. 
Now we compute the first term in the numerator of \eqref{eq:nearest_computable} as 
\begin{align}
    & \mathbb{P}\left[\bigcap_{h \in \CMcal{H}_{\sf S}} \{\hat \Phi_{\sf S}^{h}(\mathcal{A}(r,h)) = 0\}\right]  = \prod_{h \in \CMcal{H}} \mathbb{P}[\hat \Phi_{\sf S}^{h}(\mathcal{A}(r,h)) = 0] \nonumber \\
    &= \exp\left(- \frac{\lambda R^2_{\sf S}}{h_{\max}} \int_{h} \frac{{A}(r,h)}{(R_{\sf S} + h)^2} \,dh \right). \label{eq:lem_first} 
\end{align}
The second term in the numerator of \eqref{eq:nearest_computable} is
\begin{align}
    & \mathbb{P}\left[ \bigcap_{h \in \CMcal{H}} \hat \Phi_{\sf S}^{h}(\mathcal{A}(R(h),h) \backslash \mathcal{A}(r,h)) = 0 \right] \nonumber \\
    &  = \exp\left(-\frac{\lambda R^2_{\sf S}}{h_{\max}} \int_{h}  \frac{A(R(h),h)  -  {A}(r,h)}{(R_{\sf S}+h)^2} \, dh \right). \label{eq:lem_second}
\end{align}
Now we put \eqref{eq:lem_first} and \eqref{eq:lem_second} together, which leads to 
\begin{align}
    &\mathbb{P} [R > r |  \CMcal{N}] =\nonumber \\
    & \frac{\left\{ \begin{array}{l} { \exp\left(- \frac{\lambda R^2_{\sf S}}{h_{\max}} \int_{h} \frac{{A}(r,h)}{(R_{\sf S} + h)^2} \, dh \right) \cdot} \\ {\left[1- \exp\left(-\frac{\lambda R^2_{\sf S}}{h_{\max}} \int_{h} \frac{A(R(h),h)  -  {A}(r,h)}{(R_{\sf S}+h)^2} \, dh \right) \right]} \end{array}\right\}}{1- \exp\left( -\frac{\lambda R_{\sf S}^2}{h_{\max}} \int_{0}^{h_{\max}} \frac{A(R(h),h)}{(R_{\sf S} + h)^2} \, dh \right)}.
\end{align}
We finally derive the conditional PDF. Since the conditional PDF is obtained by taking derivative to the conditional CDF regarding $r$, we have
\begin{align}
    &f_{R| \CMcal{N}}(r) = \frac{\partial F_{R |\CMcal{N}}(r)}{\partial r}   \nonumber \\
    &= \frac{\frac{2 \pi r \lambda R_{\sf S}^2}{h_{\max} R_{\sf E} } \int_0^{h(r)} \frac{1}{R_{\sf S}+h}\,dh\,\cdot   \exp\left(- \frac{\lambda R^2_{\sf S}}{h_{\max}} \int_{0}^{h(r)} \frac{{A}(r,h)}{(R_{\sf S} + h)^2} \, dh \right) }{1- \exp\left( -\frac{\lambda R_{\sf S}^2}{h_{\max}} \int_{0}^{h_{\max}} \frac{A(R(h),h)}{(R_{\sf S} + h)^2} \, dh \right)}, \label{eq:lem_nearest_pdf}
\end{align}
We define $h(r) = \min(r-h_{\sf S},h_{\text{max}})$. 
This is because if $r < h_{\sf S} + h_{\max}$, the probability that the satellite is located within the spherical cap $\mathcal{A}(R(h),h)$ for $h>r-h_{\sf S}$ is equal to $0$. Otherwise, we need to calculate for all $h \in \CMcal{H}$.
This completes the proof.

\section{Proof of Lemma \ref{lem:int_laplace} } \label{proof:laplace}
Recall that $\CMcal{I} = \sum_{{\bf \hat{d}}_i \in \hat{\Phi}_{\sf I}} \bar G X_i\|\hat {\bf d}_{i}-{\bf u}_1\|^{-\beta}$. Conditioned on that the distance from the closest satellites is $r$, the conditioned interference is denoted by $\CMcal{I}_r$. 
Since the SPPP is partitioned by the altitude index $h$ as in \eqref{eq:Phi h}, the set of interfering satellites can be expressed as a union of the interfering sets at each altitude:
\begin{align}
    \hat{\Phi}_{\sf I_r} = \bigcup_{h \in \CMcal{H}} \hat{\Phi}_{\sf I_r}^h,
\end{align}
where $\hat{\Phi}_{\sf I_r}^h = \{\hat{\mathbf{d}}_i | \|\hat{\mathbf{d}}_i -\mathbf{u}_1\| > r \text{ for } \hat{\mathbf{d}}_i \in \hat \Phi^h_{\sf S} \}$. This definition signifies that the interfering satellites are located at a distance greater than $r$ from the typical user. Therefore, for each altitude $h$, the corresponding interference region is the visible region at that altitude, with a circular void of radius $r$ centered on the user. As illustrated in Fig.~\ref{fig:system_model} (d), this region is defined as:
\begin{align}
    \mathcal{A}^{\text{c}}(r,h) = \mathcal{A}(R(h), h) \setminus \mathcal{A}(r, h),
\end{align}
where $\mathcal{A}(R(h), h)$ denotes the visible region (i.e., spherical cap) at altitude $h$, and $\mathcal{A}(r, h)$ is the region within a distance $r$ from the typical user.
This decomposition is valid since $\mathcal{A}(r,h_1)$ and $\mathcal{A}(r,h_2)$ are independent for $h_1 \ne h_2$, due to the independent thinning property of the PPP.
Then, we decompose the interference according to $h \in \CMcal{H}$ such as
\begin{align}
    \CMcal{I}_r = \sum_{h \in \CMcal{H}} \CMcal{I}_r^h = \sum_{h \in \CMcal H} \sum_{{\bf \hat{d}}_i \in \hat{\Phi}_{\sf I}^h} \bar G X_i\|\hat {\bf d}_{i}-{\bf u}_1\|^{-\beta}
\end{align}
By using this, the conditional Laplace transform can be represented by 
\begin{align}
    & \CMcal{L}_{\CMcal{I}_r|\CMcal{N}}(s) = \mathbb{E}\left[e^{-s\CMcal{I}_r } \mid \CMcal{N} \right]
     = \prod_{h \in \CMcal{H}}\mathbb{E}\left[  e^{-s \CMcal{I}_r^h} \mid \CMcal{N} \right].
\end{align}
Then we have 
\begin{align}
    &\mathbb{E}\left[  e^{-s \CMcal{I}^h} \mid \CMcal{N} \right] \nonumber \\
    &\mathop{=}^{\text{(a)}} \exp\left(-\Delta \lambda(h) \int_{v \in \mathcal{A}^{c}(r,h) } \left( 1 - \mathbb{E}_X \left[ e^{-s \tilde G  X {v}^{-\beta}}\right] \right) dv \right), \nonumber \\
    &\mathop{=}^{\text{(b)}} \exp\left(-\Delta \lambda(h)  \int_{v\in \mathcal{A}^{c}(r,h)}  \left(1 - \frac{1}{(1 + s \eta \tilde{G}  v^{-\beta})^{\alpha}} \right)dv \right), \nonumber \\
    &\mathop{=}^{\text{(c)}} \exp \left( -\frac{2\pi\lambda R^2_{\sf S} \Delta h}{h_{\max} R_{\sf E} }  \int_{\max(r,h_{\sf S}+h)}^{R(h)} \left(1 - \frac{1}{(1 + s \eta \tilde{G} v^{-\beta})^{\alpha}} \right) v\, dv \right), \label{eq:laplace given h}
\end{align}
where (a) comes from the probability generating functional (PGFL) of a PPP. (b) follows the moment generating function (MGF) of the gamma distribution. 
(c) is from \eqref{eq:lambda}. In (c), the integration starts form $\max(r,h_{\sf S}+h)$ because the distance of all satellites in some spherical cap is greater than $r$. Thus, the integration region is determined according to $r$ and $h$. By combining all $h \in \CMcal{H}$ with \eqref{eq:laplace given h}, we obtain 
\begin{align}
   &\CMcal{L}_{\CMcal{I}_r|\CMcal{N}}(s) = \exp \left( - \frac{2\pi\lambda R^2_{\sf S}}{h_{\max} R_{\sf E}} \int_0^{h_{\sf max}} \frac{1}{R_{\sf S}+h}  \right. \nonumber \\
    &\quad \times \int_{\max(r,h_{\sf S}+h)}^{R(h)} \left. \left(1 - \frac{1}{(1 + s \eta \bar{G}  v^{-\beta})^{\alpha}} \right) v \, dv \, dh \right). \nonumber
\end{align}
This completes the proof.

\section{Proof of Theorem \ref{thm:coverage prob} } \label{proof:thm:coveage prob}

We recall that the rate coverage probability is expressed as 
\begin{align}
         P_{\sf cov}^{\text{SINR}}(\tau, \lambda, f_{H}, R_{\sf S}) = \mathbb{P}[\CMcal{N}] \mathbb{P}[\text{SINR} > \tau |\CMcal{N}].
\end{align}
Note that the conditional rate coverage probability is 
\begin{align}
    &\mathbb{P}\left[ \text{SINR} > \tau | \CMcal{N} \right]  = \mathbb{E}\left[ \mathbb{P}\left[ {\text{SINR}} >  \tau \middle| \CMcal{N}, \|\hat {\bf d}_{1}-{\bf u}_1\|=r \right] \right],  \nonumber \\
    &\mathop{=}^{\text{(a)}} \mathbb{E}\left[\mathbb{P}_{|\CMcal{N}} \left[ X_1 > r^{\beta}\tau\left( \CMcal{I}_r + \sigma^2 \right) \right] \right] \label{eq: p cov origin}
    \\
    &\mathop{=}^{\text{(b)}} \mathbb{E}\left[\sum_{k = 0}^{\alpha-1} \frac{1}{k!}\left( \frac{r^\beta \tau  U_r}{\eta} \right)^k e^{-\frac{r^\beta \tau U_r}{\eta}} \right] \nonumber \\
    &\mathop{=}^{\text{(c)}} \mathbb{E}_r \left[ \sum_{k = 0}^{\alpha-1} \frac{1}{k!}\left( \frac{r^\beta \tau }{\eta} \right)^k (-1)^k \frac{\partial^k \CMcal{L}_{U_r}(s) }{\partial s^k} \Bigl|_{s = \frac{r^\beta \tau}{\eta} } \right] \nonumber \\
    &\mathop{=}^{\text{(d)}} \int^{R(h_{\sf max})}_{h_{\sf S}} f_{R|\CMcal{N}}(r) \sum_{k = 0}^{\bar \alpha-1} \frac{1}{k!}\left( -s \right)^k \frac{\partial^k \CMcal{L}_{U_r}(s) }{\partial s^k} \Bigl|_{s = \frac{r^\beta \tau}{\eta} } \, dr \nonumber
\end{align}
where step (a) follows from the SINR expression in~\eqref{eq:sinr}. We assume that $\alpha$ is a natural number.
(b) is from  the conditional CCDF of the Gamma distribution $\Gamma(\alpha,\eta)$, i.e.,
\begin{align}
    \mathbb{P}_{|\CMcal{N}} \left[ X > x \right] 
    &= \sum_{k = 0}^{\alpha - 1} \frac{1}{k!} \left( \frac{x}{\eta} \right)^k e^{ - \frac{x}{\eta} }, \label{eq:ccdf_gamma}
\end{align}
for $\alpha \in \mathbb{N}$. Then, by letting $U_r = \CMcal{I}_r + \sigma^2$. This justifies step (b):
\begin{align}
    \mathbb{P}\left[ \text{SINR} > \tau \,\middle|\, \CMcal{N} \right]
    &= \mathbb{E}_r \left[ \sum_{k = 0}^{\alpha - 1} \frac{1}{k!} \left( \frac{r^\beta \tau U}{\eta} \right)^k e^{ - \frac{r^\beta \tau U}{\eta} } \right].
\end{align}
Step (c) applies the derivative identity for Laplace transforms, by applying $\mathbb{E}_{U_r} [{U_r}^k e^{-sU_r}] = (-1)^k \frac{\partial^k \CMcal{L}_{U_r}(s)}{\partial s^k}$. (d) resolves the outer expectation with respect to the randomness in the link distance $r$:
\begin{align}
    &\mathbb{P}\left[ \text{SINR} > \tau \,\middle|\, \CMcal{N} \right] \nonumber \\
    &= \int_{h_{\sf S}}^{R(h_{\sf max})} f_{R|\CMcal{N}}(r) \sum_{k = 0}^{ \alpha - 1} \frac{1}{k!} (-s)^k \frac{\partial^k \CMcal{L}_{U_r}(s)}{\partial s^k} \Bigg|_{s = \frac{r^\beta \tau}{\eta}} dr. \label{eq:Pcov given N}
\end{align}
By substituting $\CMcal{L}_{U_r}(s) = e^{-s \sigma^2} \CMcal{L}_{\CMcal{I}_r}(s)$ and applying \eqref{eq:Pcov given N} along with $\mathbb{P}[\CMcal{N}]$ from Lemma~\ref{lem:vis_prob_satellite}, the proof is complete.

\section{Proof of Corollary \ref{coro: p cov gam approx}} \label{proof: p cov gam approx}

To obtain the coverage probability with tight bound, we adopt the following Lemma \cite{alzer:mathcomp:1997, andrews:tcom:2017}:
\begin{lemma}\label{lem:alzer}
    For a normalized gamma random variable $X$ with parameter $\alpha$, the CCDF of X can be tightly upper bounded as
    \begin{align}
        \mathbb{P}[X > x] \le 1 - \left( 1- e^{-\mu x} \right)^{\alpha} = \sum_{k=1}^{\alpha} (-1)^{k+1} \binom{\alpha}{k} e^{-\mu k x}, \nonumber
    \end{align}
    where $\mu =(\alpha!)^{-\frac{1}{\alpha}}$ and equality holds for $\alpha = 1$.
\end{lemma}
Build on this, the coverage probability with tight bound is given by
\begin{align}
    &\bar{\mathcal{P}}_{cov} = \mathbb{P}[\CMcal{N}] \mathbb{P}[\text{SINR}>\tau \mid \CMcal{N}] \nonumber \\
    &=\mathbb{P}[\CMcal{N}]\, \mathbb{E}\left[\mathbb{P}_{|\CMcal{N}} \left[ X_1 > r^{\beta}\tau\left( \CMcal{I} + \sigma^2 \right) \right] \right] \nonumber \\
    &\mathop{\approx}^{\text{(a)}} \mathbb{P}[\CMcal{N}] \, \mathbb{E}\left[ 1 - \left(1-\exp \left( -\frac{\mu r^\beta \tau U}{\eta}\right) \right)^\alpha \right] \nonumber \\
    &\mathop{=}^{\text{(b)}} \mathbb{P}[\CMcal{N}] \, \mathbb{E} \left[ \sum^{\bar \alpha}_{k=1} \binom{\bar \alpha}{k} (-1)^k \exp \left( -\frac{k \mu r^\beta \tau U}{\eta} \right) \right] \nonumber \\
    &\mathop{=}^{\text{(c)}} \mathbb{P}[\CMcal{N}] \, \mathbb{E}_r \left[ \sum^{\bar \alpha}_{k=1} \binom{\bar \alpha}{k} (-1)^k \CMcal{L}_U\left( \frac{k \mu r^\beta \tau}{\eta} \right) \right] \nonumber \\
    &= \mathbb{P}[\CMcal{N}] \, \int^{R(h_{\max})}_{h_{\sf S}} f_{R|\CMcal{N}}(r) \sum^{\bar \alpha}_{k=1} \binom{\bar \alpha}{k} (-1)^k \CMcal{L}_U\left( \frac{k \mu r^\beta \tau}{\eta} \right) \, dr \nonumber
\end{align}
where (a) comes from the tight bound in Proposition \ref{prop:tight alzer bound}. (b) is from Lemma \ref{lem:alzer}. For (c), $\CMcal{L}_U(s) = e^{-s \sigma^2} \CMcal{L}_{\CMcal{I}}(s)$. This completes the proof.

\bibliographystyle{IEEEtran}
\bibliography{RHM}

\begin{thebibliography}{10}
\providecommand{\url}[1]{#1}
\csname url@samestyle\endcsname
\providecommand{\newblock}{\relax}
\providecommand{\bibinfo}[2]{#2}
\providecommand{\BIBentrySTDinterwordspacing}{\spaceskip=0pt\relax}
\providecommand{\BIBentryALTinterwordstretchfactor}{4}
\providecommand{\BIBentryALTinterwordspacing}{\spaceskip=\fontdimen2\font plus
\BIBentryALTinterwordstretchfactor\fontdimen3\font minus \fontdimen4\font\relax}
\providecommand{\BIBforeignlanguage}[2]{{%
\expandafter\ifx\csname l@#1\endcsname\relax
\typeout{** WARNING: IEEEtran.bst: No hyphenation pattern has been}%
\typeout{** loaded for the language `#1'. Using the pattern for}%
\typeout{** the default language instead.}%
\else
\language=\csname l@#1\endcsname
\fi
#2}}
\providecommand{\BIBdecl}{\relax}
\BIBdecl

\bibitem{liu:commmag:21}
S.~Liu, Z.~Gao, Y.~Wu, D.~W. Kwan~Ng, X.~Gao, K.-K. Wong, S.~Chatzinotas, and B.~Ottersten, ``{LEO} satellite constellations for {5G} and beyond: {How} will they reshape vertical domains?'' \emph{IEEE Commun. Mag.}, vol.~59, no.~7, pp. 30--36, 2021.

\bibitem{kim:twc:25}
S.~Kim, J.~Choi, W.~Shin, N.~Lee, and J.~Park, ``Multibeam satellite communications with massive {MIMO}: {Asymptotic} performance analysis and design insights,'' \emph{IEEE Trans. Wireless Commun.}, pp. 1--1, 2025.

\bibitem{mcdowell:aas:20}
J.~C. McDowell, ``The low earth orbit satellite population and impacts of the {SpaceX Starlink} constellation,'' \emph{The Astrophysical Journal Letters}, vol. 892, no.~2, p. L36, apr 2020.

\bibitem{spacex:fcc:22}
\BIBentryALTinterwordspacing
{Federal Communications Commission}, ``{Space Exploration Holdings, LLC Request for Orbital Deployment and Operating Authority for the SpaceX Gen2 NGSO Satellite System},'' Dec 2022. [Online]. Available: \url{https://www.fcc.gov/document/fcc-partially-grants-spacex-gen2-broadband-satellite-application}
\BIBentrySTDinterwordspacing

\bibitem{spacex:fcc:24}
\BIBentryALTinterwordspacing
------, ``{Space Exploration Holdings, LLC Request for Deployment and Operating Authority for the SpaceX Gen2 NGSO Satellite System},'' Nov 2024. [Online]. Available: \url{https://www.fcc.gov/document/partial-grant-spacex-gen2-application-allow-e-band-operations}
\BIBentrySTDinterwordspacing

\bibitem{okati:tcom:20}
N.~Okati, T.~Riihonen, D.~Korpi, I.~Angervuori, and R.~Wichman, ``Downlink coverage and rate analysis of low {Earth} orbit satellite constellations using stochastic geometry,'' \emph{IEEE Trans. Commun.}, vol.~68, no.~8, pp. 5120--5134, 2020.

\bibitem{park:twc:22}
J.~Park, J.~Choi, and N.~Lee, ``A tractable approach to coverage analysis in downlink satellite networks,'' \emph{IEEE Trans. Wireless Commun.}, vol.~22, no.~2, pp. 793--807, 2023.

\bibitem{choi:tcom:2025}
C.-S. Choi and F.~Baccelli, ``A novel analytical model for {LEO} and {MEO} satellite networks based on {Cox} point processes,'' \emph{IEEE Trans. Commun.}, vol.~73, no.~4, pp. 2265--2279, 2025.

\bibitem{baccelli:book:09}
F.~Baccelli and B.~Blaszczyszyn, ``Stochastic geometry and wireless networks: {Volume} {\uppercase\expandafter{\romannumeral 1}} theory,'' \emph{Found. Trends in Networking}, vol.~3, no. 3–4, p. 249–449, Mar. 2009.

\bibitem{andrews:tcom:11}
J.~G. Andrews, F.~Baccelli, and R.~K. Ganti, ``A tractable approach to coverage and rate in cellular networks,'' \emph{IEEE Trans. Commun.}, vol.~59, no.~11, pp. 3122--3134, 2011.

\bibitem{park:twc:16}
J.~{Park}, N.~{Lee}, J.~G. {Andrews}, and R.~W. {Heath}, ``On the optimal feedback rate in interference-limited multi-antenna cellular systems,'' \emph{IEEE Trans. Wireless Commun.}, vol.~15, no.~8, pp. 5748--5762, 2016.

\bibitem{dhillon:jsac:12}
H.~S. Dhillon, R.~K. Ganti, F.~Baccelli, and J.~G. Andrews, ``Modeling and analysis of {K}-tier downlink heterogeneous cellular networks,'' \emph{IEEE J. Sel. Areas Commun.}, vol.~30, no.~3, pp. 550--560, 2012.

\bibitem{renzo:twc:15}
M.~Di~Renzo, ``Stochastic geometry modeling and analysis of multi-tier millimeter wave cellular networks,'' \emph{IEEE Trans. Wireless Commun.}, vol.~14, no.~9, pp. 5038--5057, 2015.

\bibitem{park:twc:18}
J.~Park, N.~Lee, and R.~W. Heath, ``Feedback design for multi-antenna $k$ -tier heterogeneous downlink cellular networks,'' \emph{IEEE Trans. Wireless Commun.}, vol.~17, no.~6, pp. 3861--3876, 2018.

\bibitem{park:tccn:18}
J.~Park, J.~G. Andrews, and R.~W. Heath, ``Inter-operator base station coordination in spectrum-shared millimeter wave cellular networks,'' \emph{IEEE Trans. Cogn. Commun. Netw.}, vol.~4, no.~3, pp. 513--528, 2018.

\bibitem{tian:twc:22}
Y.~Tian, G.~Pan, M.~A. Kishk, and M.-S. Alouini, ``Stochastic analysis of cooperative satellite-{UAV} communications,'' \emph{IEEE Trans. Wireless Commun.}, vol.~21, no.~6, pp. 3570--3586, 2022.

\bibitem{hou:tcom:2019}
T.~Hou, Y.~Liu, Z.~Song, X.~Sun, and Y.~Chen, ``Multiple antenna aided {NOMA} in {UAV} networks: {A} stochastic geometry approach,'' \emph{IEEE Trans. Commun.}, vol.~67, no.~2, pp. 1031--1044, 2019.

\bibitem{lahmeri:commlett:2020}
M.-A. Lahmeri, M.~A. Kishk, and M.-S. Alouini, ``Stochastic geometry-based analysis of airborne base stations with laser-powered {UAV}s,'' \emph{IEEE Commun. Lett.}, vol.~24, no.~1, pp. 173--177, 2020.

\bibitem{hayajneh:access:2018}
A.~M. Hayajneh, S.~A.~R. Zaidi, D.~C. McLernon, M.~Di~Renzo, and M.~Ghogho, ``Performance analysis of {UAV} enabled disaster recovery networks: A stochastic geometric framework based on cluster processes,'' \emph{IEEE Access}, vol.~6, pp. 26\,215--26\,230, 2018.

\bibitem{wang:surv:2025}
R.~Wang, M.~A. Kishk, and M.-S. Alouini, ``Modeling and analysis of non-terrestrial networks by spherical stochastic geometry: A survey,'' \emph{IEEE Commun. Surveys \& Tutorials}, pp. 1--1, 2025.

\bibitem{talgat:commlett:20}
A.~Talgat, M.~A. Kishk, and M.-S. Alouini, ``Nearest neighbor and contact distance distribution for {Binomial} point process on spherical surfaces,'' \emph{IEEE Commun. Lett.}, vol.~24, no.~12, pp. 2659--2663, 2020.

\bibitem{okati:pimrc31:2020}
N.~Okati and T.~Riihonen, ``Stochastic analysis of satellite broadband by mega-constellations with inclined leos,'' in \emph{Proc. IEEE 31st Annu. Int. Symp. on Pers., Indoor and Mobile Radio Commun.}, 2020, pp. 1--6.

\bibitem{okati:commlett:2023}
------, ``Stochastic coverage analysis for multi-altitude leo satellite networks,'' \emph{IEEE Commun. Lett.}, vol.~27, no.~12, pp. 3305--3309, 2023.

\bibitem{ma:pimrc35:2024}
J.~Ma, Z.~Hu, Y.~Zhang, Z.~Lu, and X.~Wen, ``Coverage analysis under multi-altitude orbits for multi-layer low earth orbit satellite constellations using stochastic geometry,'' in \emph{Proc. IEEE 31st Annu. Int. Symp. on Pers., Indoor and Mobile Radio Commun.}, 2024, pp. 1--6.

\bibitem{hourani:wcl:21}
A.~Al-Hourani, ``An analytic approach for modeling the coverage performance of dense satellite networks,'' \emph{IEEE Wireless Commun. Lett.}, vol.~10, no.~4, pp. 897--901, 2021.

\bibitem{homssi:commlett:2021}
B.~A. Homssi and A.~Al-Hourani, ``Modeling uplink coverage performance in hybrid satellite-terrestrial networks,'' \emph{IEEE Commun. Lett.}, vol.~25, no.~10, pp. 3239--3243, 2021.

\bibitem{choi:tvt:2024}
C.-S. Choi and F.~Baccelli, ``Cox point processes for multi altitude leo satellite networks,'' \emph{IEEE Trans. Veh. Technol.}, vol.~73, no.~10, pp. 15\,916--15\,921, 2024.

\bibitem{lee:access:2024}
J.~Lee, S.~Noh, S.~Jung, and N.~Lee, ``Coverage analysis of {LEO} satellite downlink networks: {Orbital} geometry dependent approach,'' \emph{IEEE Access}, vol.~12, pp. 196\,939--196\,953, 2024.

\bibitem{jung:tcom:22}
D.-H. Jung, J.-G. Ryu, W.-J. Byun, and J.~Choi, ``Performance analysis of satellite communication system under the shadowed-{Rician} fading: {A} stochastic geometry approach,'' \emph{IEEE Trans. Commun.}, vol.~70, no.~4, pp. 2707--2721, 2022.

\bibitem{bhatnagar:commlett:2014}
M.~R. Bhatnagar and A.~M.K., ``On the closed-form performance analysis of maximal ratio combining in shadowed-rician fading lms channels,'' \emph{IEEE Communications Letters}, vol.~18, no.~1, pp. 54--57, 2014.

\bibitem{bai:commmag:14}
T.~Bai, A.~Alkhateeb, and R.~W. Heath, ``Coverage and capacity of millimeter-wave cellular networks,'' \emph{IEEE Commun. Mag.}, vol.~52, no.~9, pp. 70--77, 2014.

\bibitem{talga:taes:2024}
A.~Talgat, M.~A. Kishk, and M.-S. Alouini, ``Stochastic geometry-based uplink performance analysis of iot over leo satellite communication,'' \emph{IEEE Trans. Aerosp. Electron. Syst}, vol.~60, no.~4, pp. 4198--4213, 2024.

\bibitem{abdi:twc:03}
A.~Abdi, W.~Lau, M.-S. Alouini, and M.~Kaveh, ``A new simple model for land mobile satellite channels: first- and second-order statistics,'' \emph{IEEE Trans. Wireless Commun.}, vol.~2, no.~3, pp. 519--528, 2003.

\bibitem{alzer:mathcomp:1997}
H.~Alzer, ``On some inequalities for the incomplete gamma function,'' \emph{Math. Comput.}, vol.~66, no. 218, pp. 771--778, 1997.

\bibitem{Hui:iotj:2025}
M.~Hui, S.~Zhai, D.~Wang, T.~Hui, W.~Wang, P.~Du, and F.~Gong, ``A review of {LEO}-satellite communication payloads for integrated communication, navigation, and remote sensing: Opportunities, challenges, future directions,'' \emph{IEEE Internet of Things J.}, vol.~12, no.~12, pp. 18\,954--18\,992, 2025.

\bibitem{jia:iotj:22}
H.~Jia, Z.~Ni, C.~Jiang, L.~Kuang, and J.~Lu, ``Uplink interference and performance analysis for megasatellite constellation,'' \emph{IEEE Internet of Things J.}, vol.~9, no.~6, pp. 4318--4329, 2022.

\bibitem{andrews:tcom:2017}
J.~G. Andrews, T.~Bai, M.~N. Kulkarni, A.~Alkhateeb, A.~K. Gupta, and R.~W. Heath, ``Modeling and analyzing millimeter wave cellular systems,'' \emph{IEEE Trans. Commun.}, vol.~65, no.~1, pp. 403--430, 2017.

\end{thebibliography}

\end{document}